\documentclass[manuscript,screen,nonacm]{acmart}

\AtBeginDocument{%
  \providecommand\BibTeX{{%
    \normalfont B\kern-0.5em{\scshape i\kern-0.25em b}\kern-0.8em\TeX}}}

\usepackage[utf8]{inputenc}

\usepackage{url}
\usepackage{multirow}

\usepackage{microtype}
\usepackage{subfigure}
\usepackage{booktabs} 
\usepackage{amsmath}
\usepackage{amsthm}
\usepackage{amsfonts}       
\usepackage{mathtools}
\usepackage{nicefrac}
\usepackage{algorithm,algorithmic}
\usepackage{listings}
\usepackage{balance}

\usepackage{enumerate}

\usepackage{latexsym}

\newcommand{\commentout}[1]{}

\newcommand{\R}{\mathbb{R}}

\newcommand{\bfT}{\mathbf{T}}

\newcommand{\bfv}{\mathbf{v}}
\newcommand{\bfw}{\mathbf{w}}

\newcommand{\bfR}{\mathbf{R}}

\newcommand{\calU}{\mathcal{U}}
\newcommand{\calR}{\mathcal{R}}
\newcommand{\calI}{\mathcal{I}}
\newcommand{\calT}{\mathcal{T}}
\newcommand{\calL}{\mathcal{L}}
\newcommand{\calN}{\mathcal{N}}

\newcommand{\hatr}{\hat{r}}
\newcommand{\olg}{\overline{g}}

\DeclareMathOperator*{\argmax}{argmax}

\newcommand{\veps}{\varepsilon}

\newcommand{\NumR}{\mathit{Num}}
\newcommand{\Rated}{\mathit{Rated}}
\newcommand{\Tagged}{\mathit{Tagged}}
\newcommand{\PT}{\mathit{PT}}
\newcommand{\obj}{\mathit{obj}}
\newcommand{\subj}{\mathit{subj}}

\newcommand{\UAU}{\mathit{UAU}}
\newcommand{\UMU}{\mathit{UMU}}

\newtheoremstyle{TheoremNum}%
    {\topsep}{\topsep}
    {\itshape}
    {}
    {\bfseries}
    {.}
    { }
    {\thmname{#1}\thmnote{ \bfseries #3}}
\theoremstyle{TheoremNum}

\title[Personalized Soft Attribute Semantics in Recommender Systems Using CAVs]{Discovering Personalized Semantics for Soft Attributes in Recommender Systems using Concept Activation Vectors}
\titlenote{This is an extended version of a paper that appeared at WWW-22: The Web Conference 2022, April 2022, Lyon, France.}

\author[C.~G\"{o}pfert]{Christina G\"{o}pfert}
\affiliation{%
\institution{Amazon}
  \country{Germany}
}
\email{chgopfert@gmail.com}

\author[A.~Haig]{Alex Haig}
\authornote{Contact author.}
\affiliation{%
  \institution{Google Research}
  \streetaddress{1600 Amphitheatre Parkway}
  \city{Mountain View}
  \state{CA}
  \country{USA}
  \postcode{94043}
}
\email{ahaig@google.com}

\author[C.~Hsu]{Chih-wei Hsu}
\authornote{Contact author.}
\affiliation{%
  \institution{Google Research}
  \streetaddress{1600 Amphitheatre Parkway}
  \city{Mountain View}
  \state{CA}
  \country{USA}
  \postcode{94043}
}
\email{cwhsu@google.com}

\author[Y.~Chow]{Yinlam Chow}
\authornote{Contact author.}
\affiliation{%
  \institution{Google Research}
  \streetaddress{1600 Amphitheatre Parkway}
  \city{Mountain View}
  \state{CA}
  \country{USA}
  \postcode{94043}
}
\email{yinlamchow@google.com}

\author[I.~Vendrov]{Ivan Vendrov}
\affiliation{%
  \institution{Anthropic}
  \city{San Francisco}
  \state{CA}
  \country{USA}
}
\email{ivendrov@gmail.com}

\author[T.~Lu]{Tyler Lu}
\authornotemark[2]
\affiliation{%
  \institution{Meta AI}
  \city{Menlo Park}
  \state{CA}
  \country{USA}
}
\email{tyler.lu@gmail.com}

\author[D.~Ramachandran]{Deepak Ramachandran}
\affiliation{%
  \institution{Google Research}
  \streetaddress{1600 Amphitheatre Parkway}
  \city{Mountain View}
  \state{CA}
  \country{USA}
  \postcode{94043}
}
\email{ramachandrand@google.com}

\author[H.~Pham]{Hubert Pham}
\affiliation{%
  \institution{Google Research}
  \streetaddress{1600 Amphitheatre Parkway}
  \city{Mountain View}
  \state{CA}
  \country{USA}
  \postcode{94043}
}
\email{hubertpham@google.com}

\author[M. Ghavamzadeh]{Mohammad Ghavamzadeh}
\affiliation{%
  \institution{Google Research}
  \streetaddress{1600 Amphitheatre Parkway}
  \city{Mountain View}
  \state{CA}
  \country{USA}
  \postcode{94043}
}
\email{ghavamza@google.com}

\author[C.~Boutilier]{Craig Boutilier}
\affiliation{%
  \institution{Google Research}
  \streetaddress{1600 Amphitheatre Parkway}
  \city{Mountain View}
  \state{CA}
  \country{USA}
  \postcode{94043}
}
\email{cboutilier@google.com}
\renewcommand{\shortauthors}{C. G\"{o}pfert et al.}

\begin{document}

\begin{abstract}
Interactive \emph{recommender systems} have emerged as a promising paradigm to overcome the limitations of the primitive user feedback used by traditional recommender systems (e.g., clicks, item consumption, ratings). They allow users to express intent, preferences, constraints, and contexts in a richer fashion, often using natural language (including faceted search and dialogue). 
Yet more research is needed to find the most effective ways to use this feedback. One challenge is \emph{inferring a user's semantic intent} 
from the open-ended terms or attributes often used to describe a desired item, 
and using it to refine recommendation results.
Leveraging \emph{concept activation vectors (CAVs)} \cite{kimTCAV:icml18},
a recently developed approach for model interpretability in machine learning,
we develop a framework to learn a representation that captures the semantics of such attributes and connects them to user preferences and behaviors in recommender systems. One novel feature of our approach is its ability to distinguish objective and \emph{subjective} attributes (both subjectivity of \emph{degree} and of \emph{sense}), and associate \emph{different senses} of subjective attributes with different users. 
We demonstrate on both synthetic and real-world data sets that our CAV representation not only accurately interprets users' subjective semantics, but can also be used to improve recommendations through \emph{interactive item critiquing}.
\end{abstract}




\maketitle

\section{Introduction}
\label{sec:intro}

The ubiquity of \emph{recommender systems} in mediating the discovery and consumption of content, products and services has not only driven user demand for recommender systems, but has increased the expectation that such systems should more deeply understand their needs and preferences.
\emph{Conversational recommenders} \cite{wicrs2020} have emerged as a promising paradigm to meet such expectations---they can overcome the primitive user feedback admitted by traditional recommender systems (e.g., simple queries, clicks, item consumption, ratings, purchases), and allow users to express their intent, preferences, constraints and contexts in a richer fashion through the use of natural-language-based interaction (e.g., faceted search, or more open-ended dialogue), and thereby dramatically increasing the bandwidth of communication between user and the recommender system.
However, interpreting such interactions requires grounding the intended semantics of the user with respect to the recommender system's underlying model of user preferences. For example,
if a user expresses a desire for a ``funny'' movie, this must be translated into an actionable representation of her preferences over the target movie corpus.

 When the set of item attributes is well-defined and known \emph{a priori} (e.g., as with the item attributes in a product catalog), existing techniques such as \emph{faceted search} \cite{koren_facted:www08,zheng_faceted_survey:2013} or \emph{example critiquing} \cite{burke-critiquing,chen_critiquing_survey:umuai2012} can be used directly. But often item attributes are \emph{soft} \cite{sigir21:filipandkristian}: there is no definitive ``ground truth'' source associating such soft attributes with items; the attributes themselves may have imprecise interpretations; and they may be \emph{subjective} in nature (i.e., different users may interpret them differently). For instance, in \emph{collaborative filtering (CF)} tasks such as movie recommendation, side information about movie attributes like `funny,' `thought-provoking,' or `violent' is often available, but it is often
 ancillary, derived from sparse, noisy user comments, reviews, or tags. Moreover, users may disagree on which movies they consider to be `violent' (or which are `too violent' for them). Using soft attributes for item search or recommendation is thus challenging for three main reasons. First, which items exhibit such attributes usually has to be predicted, in the absence of a ground truth, using models built from noisy, incomplete data, and is often characterized by considerable uncertainty. Second, these predictions may need to depend on a specific user's usage or semantics, reflecting potential subjectivity in the semantics of such attributes. Third, their open-ended nature means that recommender systems must often take steps to \emph{identify} those soft attributes that are most relevant to their domain and to their users.
 
 Recent work has attempted to \emph{jointly} learn the semantics of soft attributes with user preferences \cite{wu2019deep,luo2020latent,nema2021untangle}, providing some steps toward the question of soft attribute semantics. 
 In this work, we adopt a different perspective: we treat the recommendation task as primary, using standard collaborative filtering models
 for recommender systems without adjusting them to incorporate soft attibutes. Instead, we \emph{infer the semantics of soft attributes using the representation learned by the recommender system model itself} \cite{rendle2010,cohen:recsys2017}.
Our approach has four key advantages over joint methods:
\begin{enumerate}[(1)]
\item The recommender system's model capacity is directed to predicting user-item preferences without further trying to predict additional side information (e.g., tags), which often does not improve recommender system performance.
\item The recommender system model can easily accommodate new attributes \emph{without retraining} should new sources of tags, keywords or phrases emerge from which to derive new soft attributes. 
\item Our approach offers a means to test whether specific soft attributes are \emph{relevant} to predicting user preferences. Thus, we are able focus attention on attributes most relevant to capturing a user's intent (e.g., when explaining recommendations, eliciting preferences, or suggesting critiques).
\item One can learn soft attribute/tag semantics with relatively small amounts of labelled data, in the spirit of pre-training and few-shot learning.
\end{enumerate}
 
At a high-level, our approach works as follows. we assume we are given: (i) a collaborative filtering-style model (e.g., probabilistic matrix factorization or dual encoder) which embeds items and users in a latent space based on user-item ratings; and (ii) a (small) set of \emph{tags} (i.e., soft attribute labels) provided by a \emph{subset of users} for a \emph{subset of items}. We develop methods that associate with each item the degree to which it exhibits a soft attribute, thus determining that attribute's semantics. We do this by applying \emph{concept activation vectors (CAVs)} \cite{kimTCAV:icml18}---a recent method developed for interpretability of machine-learned models---to the collaborative filtering model to detect whether it \emph{learned a representation of the attribute}. The projection of this CAV in embedding space provides a (local) \emph{directional semantics} for the attribute that can then be applied to items (and users). Moreover, the technique can be used to identify the \emph{subjective nature} of an attribute, specifically, whether different users have different meanings (or tag \emph{senses}) in mind when using that tag. Such a \emph{personalized semantics} for subjective attributes can be vital to the sound interpretation of a user's true intent when trying to assess her preferences.
 
Our key contributions in this work are as follows:
\begin{enumerate}[(1)]
\item We propose a novel framework using CAVs to identify the semantics of soft attributes relative to preference prediction or behavioral models in recommender systems \emph{without requiring co-training} of semantics and preference models.
\item We develop methods to distinguish \emph{objective} and \emph{subjective} attributes (both subjectivity of
\emph{degree} and of \emph{sense}) and associate different senses of subjective attributes with different users.
\item We investigate a simple method that leverages this semantics to elicit preferences via \emph{example critiquing}.
\end{enumerate}

The remainder of the paper is organized as follows. In Section~\ref{sec:formulation} we outline our problem setup and discuss related work. In Section~\ref{sec:objectiveCAVs}, we use CAVs to identify the semantics of (objective) soft attributes by leveraging collaborative filtering user/item representations (linear and nonlinear), while in Section~\ref{sec:subjectiveCAVs} we extend the approach to handle subjective attributes and identify a user's interpretation for different tags.  We test our methods on both synthetic data and the MovieLens20M data set \cite{harper16:movielens} in these two sections. In
Section~\ref{sec:user_study} we test the ability of CAVs to uncover the semantics of soft attributes relative to semantic assessments of items provided by human raters, using the \emph{SoftAttributes} data set, a rater-generated soft-attribute data set developed by \citet{sigir21:filipandkristian}. In Section~\ref{sec:critiquing} we illustrate how CAVs can be used to support example critiquing using such soft attributes. We also include appendices that describe some of the details omitted (but summarized) in the main body of the paper---this is done to improve the flow of the main text.

\section{Problem Formulation}
\label{sec:formulation}

We first outline our problem formulation and data assumptions, then briefly discuss related work.

\vskip 2mm
\noindent
\textbf{User-item Ratings.} \hspace*{2mm}
We assume a standard collaborative filtering task: users $\calU$ offer ratings of items $\calI$, with $r_{u,i}\in\calR$ denoting
the rating of user $u\in\calU$ for item $i\in\calI$, where $\calR$ is the set of possible ratings
(e.g., 1--5 stars). Let $n=|\calU|, m=|\calI|$, and $\bfR$ denote the $m\times n$ (usually sparse) ratings matrix,
with $r_{u,i}\!=\! 0$ denoting that user $u$ has not rated item $i$. Let $R = \{(u,i) : r_{u,i} \neq 0\}$ be the set of user-item pairs for which a rating has been given.

\vskip 2mm
\noindent
\textbf{Preference Predictions.} \hspace*{2mm}
We assume a collaborative filtering method has been applied to the ratings matrix $\bfR$ to construct \emph{user and item embeddings}, $\phi_U:\calU \mapsto \R^d$ and $\phi_I: \calI \mapsto \R^d$, respectively, such that the model's predicted (expected) rating for any user-item pair is $\hatr_{i,u} = \phi_U(u)^\top \phi_I(i)$. We let $X\subseteq \R^d$ generically denote the embedding space. Suitable methods for generating such embedding representations include
matrix factorization \cite{salakhutdinov-mnih:nips07} or certain forms of neural collaborative filtering
\cite{beutel_etal:wsdm18,yangEtAl:www20}. For concreteness,
we assume a \emph{two-tower model} (or \emph{dual encoder}) in which users and items are passed through separate (but co-trained) deep neural nets (DNNs), $N_U$ and $N_I$, to produce their respective vector embeddings $\phi_U(u)$ and $\phi_I(i)$, which are combined via dot product to predict user-item affinity $\hatr_{i,u}$ \cite{yiEtAl:recsys19,yangEtAl:www20}. 
We can view $\phi_I(i)$ as a (learned) \emph{latent feature vector} characterizing item $i$ and $\phi_U(u)$ as parameterizing user $u$'s estimated \emph{utility (or preference) function} over these ``latent features.'' Notice that, by construction, this interpretation means user utility is linear with respect to these latent item features. This is a limitation we discuss in some depth in Section~\ref{sec:nonlinear}.
While we focus on this specific two-tower architecture, our methods apply directly to matrix factorization models and should extend to more general 
neural collaborative filtering approaches \cite{he_etal:www17}.

\vskip 2mm
\noindent
\textbf{Soft Attributes \& Tags.} \hspace*{2mm}
Collaborative filtering methods are often used to predict user-item affinity in \emph{content} recommender systems (e.g., for movies, music, news, etc.) because \emph{user rating or consumption behavior} is generally far more predictive of user preferences than typical \emph{hard attributes}. Here we use the term to denote the commonly known and ``objective'' features or properties of (e.g., genre, artist, director) \cite{grouplens:cacm97}. Despite this, users often \emph{describe} items of interest using \emph{soft attributes} \cite{radlinski:sdd2019,sigir21:filipandkristian}, features that are not part of an agreed-upon, formal item specification. For example, movies might be described using terms like `funny,' `thought-provoking,' `violent,' `cheesy,' etc.

Often soft attributes are articulated by users via terms or \emph{tags} applied to items, which are intended to refer to some underlying attribute or property of the item in question. The use of tags makes it clear why these underlying attributes are soft: the tags are neither applied universally to all items, nor are they applied by all users. Moreover, as we discuss below, these tags may be applied \emph{subjectively} in the sense that users may disagree on the items to which, or the degree to which, these tags apply. Finally, their often open-ended nature makes it challenging to determine which soft attributes are most closely aligned with, or predictive of, user preferences. It is these three properties that require a recommender system to treat the attributes corresponding to tags as soft. 

A number of recommender systems support user-supplied tags (see, e.g., the MovieLens data set \cite{harper16:movielens} we use below). In what follows, we will make this assumption; however, tags may also be extracted from user descriptions, reviews or other data sources \cite{mining_reviews}. For simplicity, we assume a set of $k$ canonical tags $\calT$ that users may adopt to describe items.\footnote{We set aside the question of aggregating potentially diverse tags into a set of covering prototypes.}
We assume that tags are used \emph{propositionally}, that is, a user who considers some tag for some item simply chooses to apply the tag to an item or not. However, even in this propositional usage, the underlying attribute to which a tag refers may be \emph{ordinal} or \emph{cardinal}, e.g., a tag `violent' may refer to some degree of `violence'.\footnote{Our techniques can be extended in a straightforward way to Boolean (positive and negative application), ordinal or cardinal tags.} \emph{Tag data} comprises a $m\times n \times k$ tensor $\bfT$ where $t_{u,i,g} = 1$ if user $u$ applies tag $g$ to item $i$, and $0$ otherwise. Let $T = \{(u,i,g) : t_{u,i,g}\!=\! 1\}$ be the set of user-item-tag triples in the data set, and $T_g = \{(u,i) : t_{u,i,g}\!=\! 1\}$ be the set of user-item pairs for which tag $g$ has been applied. Tags are usually strictly sparser than ratings---that is, if a user $u$ tags an item $i$ with \emph{any} tag $g$, $u$ has also rated $i$---so we assume $T_g \subseteq R$ for all tags $g$. Note that user $u$ may apply multiple tags to the same item (e.g., a user may tag a movie as `violent,' `thought-provoking' and `dark'). Let $T_{u} \subseteq\calI$ be the set of items tagged by $u$ (using any tag), $T_{u,g}$ be those tagged by $u$ with $g$ specifically, and
$T_{u,\olg} = T_{u} \setminus T_{u,g}$ be those tagged by $u$ but \emph{not} with $g$. Our tag-data setup is similar to that used in work on tag recommenders \cite{gantner:icdm2010,cohen:recsys2017,luo2020latent}.

\vskip 2mm
\noindent
\textbf{Elicitation \& Critiquing.} \hspace*{2mm}
Collaborative filtering models, in isolation, are ill-suited to recommender systems that aim to naturally interact with users to refine knowledge of their preferences. A CF-based recommender system can actively elicit new ratings or support ``more like this'' statements at the \emph{item level} \cite{activecf:uai03,zhao13interactive}, but the embedding representation of items, since it is not generally interpretable, does not lend itself to such \emph{attribute-based interaction}. Tags can help alleviate this limitation, and offer a compelling mechanism to help users navigate the item space.
A number of preference elicitation and example critiquing methods have been developed that use \emph{hard attributes} (see Related Work below). The challenge in content recommenders is that tags typically correspond to \emph{soft attributes} and are often \emph{subjective} in nature. For example, if a user critiques a movie recommendation by asking for something ``more thought-provoking'' or ``less violent,'' standard techniques for adjusting the recommender system's model of the user's preference cannot be applied unless we have a concrete semantics that relates the corresponding tag---or more precisely, the underlying soft attribute---to specific items in a way that better reflects her preferences. It is this problem we address in this work.

\vskip 2mm
\noindent
\textbf{Concept Activation Vectors.} \hspace*{2mm}
Research on interpretable representations tries to overcome the fact that modern machine-learned models---and deep neural networks (DNNs) in particular---usually learn complex, non-transparent representations of concepts \cite{sundararajan:icml2017,kimTCAV:icml18}.
The \emph{testing CAVs (TCAV)} framework
\cite{kimTCAV:icml18} is one such mechanism that tries to find a correspondence between the ``state'' of a model (e.g.,
input features, DNN activation patterns) and human-interpretable concepts. For instance, suppose a DNN has been trained to classify animals in images. Using only a small set of images with positive and negative examples of some concept (e.g., ``objects with stripes''), TCAV is used to test whether the DNN has learned a representation of that concept in the form of a vector of activation (CAV) that correlates with its presence. Moreover, using the derivative of the classifiers output with respect to the CAV's direction, it measures how important that concept is to the classifier's predictions (e.g., how sensitive a ``zebra'' classification is to the presence of stripes in an image). In our setting, we use CAVs to translate between latent item representations learned by a collaborative filtering model and the soft attributes that users adopt to describe items and preferences. We provide more detail on the application of CAVs to recommender item representations in the next section.

For recommender systems, we use CAVs to translate between the latent item representations learned by a collaborative filtering model and the soft attributes users adopt to describe items and preferences. We detail the adaptation of key CAV concepts to recommender systems in Section~\ref{sec:objectiveCAVs} below.

\vskip 2mm
\noindent
\textbf{Related Work.} \hspace*{2mm}
A number of methods exist for finding the semantics of tags and attributes in recommender systems using tag data \cite{gantner:icdm2010,luo2020latent} or reviews \cite{mcauley:icdm2012}. While some learn semantics jointly with ratings prediction, others build attribute models ``on top of'' a latent factor model built for ratings prediction as we do here. Most related to our approach is that of \citet{gantner:icdm2010}, who learn semantics for tags or explicit item features as a linear combination of latent features using $k$-nearest neighbors or linear regression, where the latent features are those of a Bayesian personalized regression (BPR) model \cite{rendle_BPR:uai09} (though the technique applies more broadly). \citet{cohen:recsys2017} extend this model to incorporate uncertainty and active exploration. This line of work is proposed as a means for solving the cold-start problem. Our work differs primarily in its ability to handle nonlinear representations and subjectivity, and in its focus on conversational and critiquing recommender systems. More generally, the broader literature on tag recommendation bears some connection to our work \cite{krestel:recsys09,rendle2010,krohn-grimberghe:wsdm12,marinho:ecml09} by modeling the relationship between users, items and tags.

 One of our main motivations is to make soft and subjective attributes interpretable for use in critiquing \cite{chen_critiquing_survey:umuai2012}, faceted search \cite{zheng_faceted_survey:2013}, and preference elicitation \cite{cf-survey:2005,viappiani:06,priyogi19} for recommender systems. A wide variety of techniques have been developed in these areas, though we do not survey these here since we do not propose novel elicitation or critiquing methods---our focus is instead on how to incorporate soft, subjective attributes into such methods, which we exemplify using a fairly ``vanilla'' critiquing model. However, \citet{radlinski:sdd2019} develop a methodology for connecting the preferences of users with their use of soft attributes in conversational recommender systems, while \citet{sigir21:filipandkristian} develop techniques and data set (which we use below) to assess the semantics of soft attributes.
 
While little work in elicitation for recommendation addresses subjectivity, one exception is \cite{brv:icml09,brv:aaai10} who consider ``definitional'' subjectivity defined using personalized logical concepts, which is a very different notion from ours. \citet{aaaisubjattributes} provides an overview of various research challenges associated with the use of subjective attributes.
Subjectivity has been studied in natural language processing and psycholinguistics. \citet{welch2020} show that training personalized embeddings for certain classes of words (e.g., adverbs, social words, words relating to cognitive processes) increases language modeling performance. Their methodology could be used for a more granular analysis of which words and phrases are most subjective, which can then be used as a prior for analyzing subjectivity in recommender settings. Prototype theory \cite{osherson1981} is a popular methodology where subjectivity is related to the similarity of an item to an idealized exemplar. 
Geometrically, this could be considered analogous to a ``ball'' semantics of subjectivity while CAVs encaspulate a ``line'' semantics. 

\section{Finding Relevant Soft Attributes}
\label{sec:objectiveCAVs}

Our first contribution is the development of a novel method for identifying the semantics of \emph{relevant soft attributes} with respect to the item embedding representation learned by our collaborative filtering method.  Assume a collaborative filtering model $\Phi = (\phi_U, \phi_I)$ trained on ratings data $\bfR$ with additional tag data $\bfT$. As above, we will often assume that $\phi_U$ and $\phi_I$ are realized by DNNs $N_U$ and $N_I$, respectively. We use CAVs \cite{kimTCAV:icml18}
to discover whether the collaborative filtering model has learned an implicit representation of a soft attribute corresponding to a given tag. If so, that representation can be used to support example critiquing, preference elicitation or item-space navigation (Section~\ref{sec:critiquing}).
Critically, \emph{we do not use the tag data} when training the collaborative filtering model---this is akin to work that builds attribute models on top of embeddings to address the cold-start problem \cite{rendle2010,cohen:recsys2017}, and stands in contrast with approaches that jointly train models to predict both ratings and attributes \cite{wu2019deep,luo2020latent}). Our hypothesis is the following:
\begin{quote}
\emph{If a tag is useful for understanding user preferences across a broad swath of the user population, the collaborative filtering model will have learned a representation of the corresponding soft attribute}. 
\end{quote}
As such, our general approach takes the collaborative filtering model as primary, and works directly with it. Of course, the converse is that if no such representation (or CAV) is uncovered by our approach, the soft attribute in question is of limited use for users expressing their preferences.

Our approach has a several important advantages:
\begin{enumerate}[(1)]
\item The recommender system model can be developed/trained/used without a pre-commitment to a specific attribute vocabulary---new attributes can be added as needed without retraining the model itself.
\item Recommender system model capacity is focused on the core task of preference prediction and recommendation, and not used for attribute prediction directly.
\item Our method can be used to assess the relevance and importance of specific attributes for preference elicitation or critiquing.
\end{enumerate}
In this section, we set aside 
the possibility of
subjectivity in attribute semantics, deferring its treatment to Section~\ref{sec:subjectiveCAVs} (though some implicit \emph{subjectivity of degree} is accounted for here). We begin with an informal elaboration of how CAVs can be applied directly to soft attribute semantics in recommender systems (Section~\ref{sec:mapping}). We explicate our technical approach by first assuming that soft attributes are linear in user/item embedding space (Section~\ref{sec:linear}), then turn to nonlinear representations (Section~\ref{sec:nonlinear}). We evaluate our methods empirically in Section~\ref{sec:empiricalObjective}.

\subsection{Mapping CAV Notions to Recommender Systems}
\label{sec:mapping}

\begin{figure}
    \centering
\includegraphics[width=0.9\textwidth]{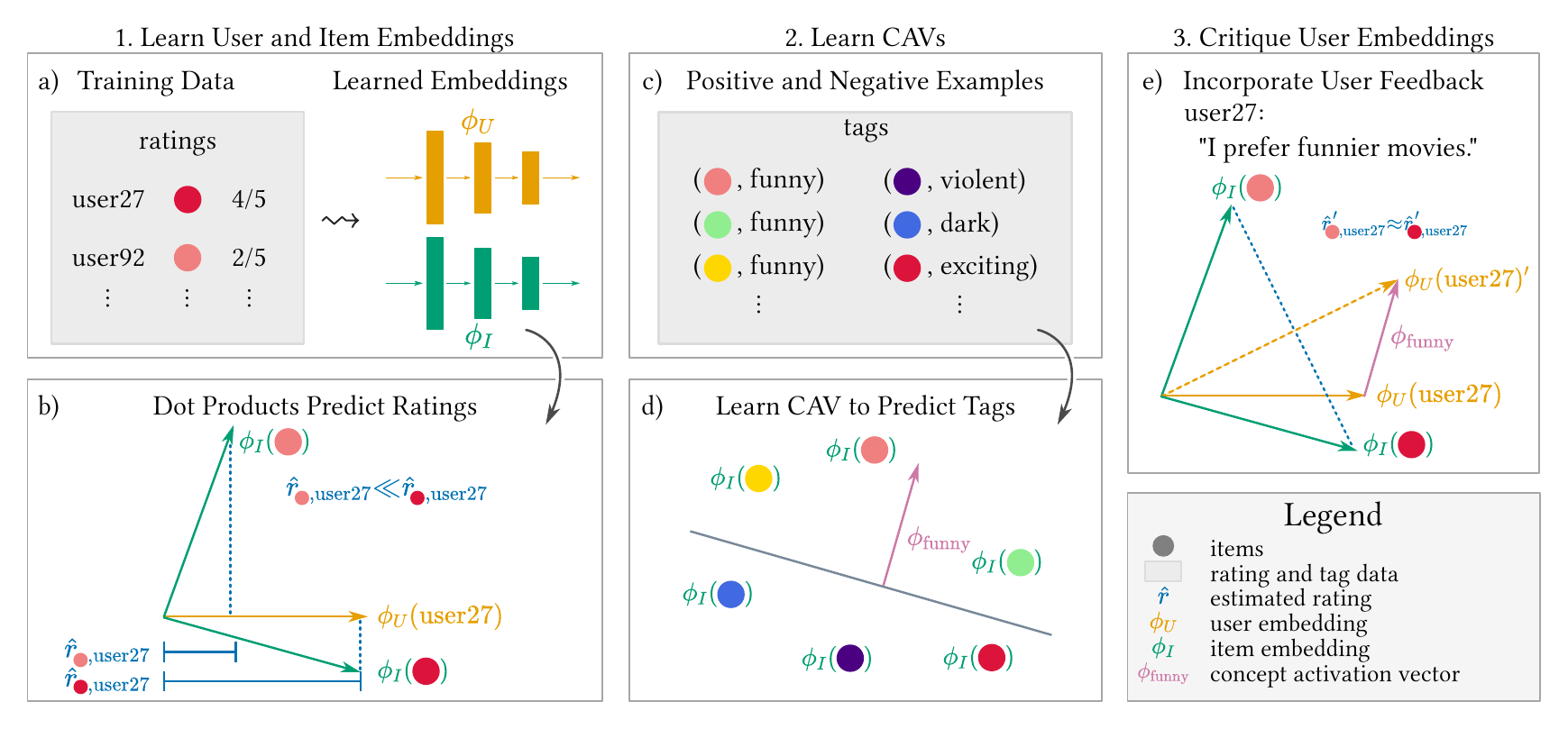}
  \vspace{-0.1in}
\caption{An overview of the collaborative filtering, CAV learning and critiquing setup used in our work. a) Learn user $\phi_U$ and item embeddings $\phi_I$ from ratings data. b) User-item affinity $\hatr_{i,u}$ is predicted using dot product of user-item embeddings. c) To learn a CAV for the concept `funny,' gather positive and negative examples, e.g., from tag data, and use them to d) learn a CAV in item-embedding space. (We explore several methods for CAV training.) e) In one use case, we use the CAV to update a user's embedding given her item-attribute feedback. Figure inspired by \cite{kimTCAV:icml18}.}
  \vspace{-0.1in}
\label{fig:overview}
\end{figure}

Before formalizing our approach, we briefly map (and adapt) key CAV concepts to our recommender setting by drawing an analogy with the image classification setting used to explicate CAVs by \citet{kimTCAV:icml18} (and briefly described above). Our DNN collaborative filtering model $\Phi = (\phi_U, \phi_I)$ is trained on user-item ratings, similar to a multi-class image classifier trained on labeled images. A soft attribute in our setting corresponding to some tag $g$ (say, `violent') is analogous to some image feature (e.g., `stripes') in the image setting. Our aim is to use CAVs to determine if the trained \emph{item network} $N_I$ has learned some representation of the attribute corresponding to this tag, by direct analogy to the image case, where the goal is to uncover a representation of an image feature like `stripes' if one exists in the classifier network. Just as in the image setting, where some small set of positive example images (with stripes) and negative examples (non-striped) is used to attempt to identify a CAV, we use a \emph{small} set of positive (tagged) and negative (untagged) items for the same purpose, though we must account for both variability and inconsistency in the tags applied by different users.\footnote{While not our aim in this work, we can also use CAVs to test the sensitivity of various predictions to the presence of this soft attribute: by analogy with testing the sensitivity of a `zebra' classification to the presence of stripes, we can test the sensitivity of a user's item rating to the item's (degree of), say, violence. In the collaborative filtering setting however, this sensitivity will differ across the user population.}
Fig.~\ref{fig:overview} offers a graphical depiction of our use of CAVs for recommender systems (including in example critiquing, see Section~\ref{sec:critiquing}).
We refer to 
Table~\ref{tab:keyconcepts} for a concise list of key concepts and notations.

\begin{table}
    \centering
    \begin{tabular}{|p{2.6cm}|p{12cm}|}
    \hline
       \emph{Rating}  &  Measure $r_{u,i}$ of user $u$'s preference for item $i$.\\ \hline
       \emph{User/item embedding}  &  Vector representations of users $\phi_U(u)$ and items $\phi_I(i)$ learned using, for example, collaborative filtering methods on ratings data. The estimate of $u$'s rating for $i$ is $\hatr_{i,u} = \phi_U(u)^\top \phi_I(i)$. If $\phi_I$ is represented by a DNN, we denote the activations for item $i$ at the $\ell$-th layer by $\phi_{I,\ell}(i)$.\\ \hline
       \emph{Tag}  &  A term (from set $\calT$) propositionally applied by users to describe items. Each tag corresponds to an attribute or \emph{concept}.\\ \hline
       \emph{Attribute}  &  A specific property of items (or a \emph{concept}). Attributes may be cardinal or ordinal (items can possess more or less of an attribute) or Boolean (items either exhibit that attribute or not).\\ \hline
       \emph{Concept activation vector (CAV)}  &  A CAV $\phi_g$ for a tag $g$ is a vector in embedding or activation space that represents a direction based on which items ``possess more of'' the attribute corresponding to $g$. In effect, this is the semantics of the underlying attribute. CAVs can be learned using classification or learning-to-rank methods on tag data.\\ \hline
       \emph{Subjectivity}  & We distinguish three types of tags: (1) \emph{objective} tags, where users (more or less) agree on whether (or the degree to which) an item satisfies the attribute corresponding to the tag; (2) \emph{degree subjective} tags, where users agree on the relative degree to which items possess the underlying attribute, but may disagree on whether the (boolean) attribute/tag applies; and (3) \emph{sense subjective} tags, where (groups of) users may disagree on which items possess the underlying attribute (or the relative degree).\\ \hline
    \end{tabular}
    \vspace*{2mm}
    \caption{A Recap of Several Key Concepts Used Throughout the Paper.}
    \label{tab:keyconcepts}
\end{table}

\subsection{Linear Attributes}
\label{sec:linear}

We adapt CAVs
to test whether a collaborative filtering model has learned a representation of a soft attribute corresponding to a tag. We first illustrate our approach by testing whether the \emph{embedding space itself} contains a \emph{linear} representation of the tag's underlying attribute (i.e., an attribute that is linear in item embedding features $\phi_I$).
We generalize this linear model (whose weaknesses we detail below) in Section~\ref{sec:nonlinear}.
Given collaborative filtering model $\Phi$, each item $i\in\calI$ is represented by its embedding $\phi_I(i)\in X$. For any user $u$, the representations $\phi_I(i)$ of items $i\in T_{u,g}$ to which she has applied tag $g$ are treated as positive examples of the underlying concept (say, violent movies), while the $\phi_I(i)$ for any $i\in T_{u,\olg}$ are used as negative examples. Notice that these negatives are ``implicit,'' but they are plausible---the user $u$ has otherwise tagged these items, and has not used $g$; moreover, we only consider negative examples for $g$ if $u$ has applied tag $g$ positively to at least one item (i.e., has shown an awareness of, and willingness to use, tag $g$).\footnote{Potential bias that may arise since tags may not be ``missing at random'' \cite{marlin07}, but we defer this issue to future work.}  Explicit negatives are more informative, and what follows can be generalized easily to this case. Taking examples over all $u\in\calU$, we train a linear classifier in one of two ways.

\vskip 2mm
\noindent
\textbf{Binary Logistic Regression.}\hspace*{2mm}
Our first model assumes that every user $u$ uses a tag $g$ in roughly the same way, with positive instances given by the multi-set $\cup_\calU \{i: i\in T_{u,g}\}$, and negatives given by
$\cup_\calU \{i: i\in T_{u,\olg}\}$. Since positive tag examples are often sparse, we
use \emph{negative sampling} to manage class imbalance \cite{mikolov:nips13,yangEtAl:www20}.\footnote{For example, in the MovieLens20M data set we use below, there are more than 50 times as many ratings given as there are tags used. We evaluate the ratio of positive to negative samples in Section~\ref{sec:user_study}.} Let $D_g$ be the induced ``global'' data set aggregating the data of all users. We train a logistic regressor $\phi_g$ to predict whether an item $i$ ``satisfies'' the soft attribute corresponding to tag $g$. Specifically, $P(g(i);\phi_g) = \sigma(\phi_g^\top \phi_I(i))$ is the predicted probability that $i$ satisfies $g$, and is trained using (regularized) logistic loss (and labels {\small $y \in \{+1, -1\}$}):
\begin{align}
\calL(\phi_g; D_g) =  \sum_{(i,y)\in D_g} \log(1 + e^{-y \phi_g^\top \phi_I(i)}) + \frac{\lambda}{2} \phi_g^\top \phi_g.
\label{eq:binaryloss}
\end{align}
Once trained, the regressor $\phi_g$ obtained serves as our CAV.
We refer to this method as \emph{binary logistic regression for CAVs}.

\vskip 2mm
\noindent
\textbf{Per-user Pairwise Loss (RankNet).}\hspace*{2mm}
If two users disagree on the application of tag $g$ to some item, this global classifier treats it as label noise. An alternative explanation for such a discrepancy is that they agree on the ``direction'' of $g$, but disagree on the ``degree'' to which item $i$ exhibits $g$'s underlying soft attribute. For example, given \emph{any} pair of movies, two users may agree on which movie of the pair is more violent; but they may have different \emph{thresholds} or tolerances when actually applying the tag (e.g., they might disagree on ``how violent is violent''). For instance, both users may agree that movie $i$ is more violent than movie $j$, but one may consider neither $i$ nor $j$ to reach the threshold needed for them to label it truly `violent,' while the other may believe that one or both of $i$ and $j$ warrant that label. Our second model accounts for this by treating each user $u$ as generating \emph{pairwise comparisons}, 
$$D_u = \{i \succ_g j: i\in T_{u,g}, j\in T_{u,\olg} \},$$
drawn from an underlying ranking.
Using this set of comparisons (over all $u\in\calU$, with negative sampling),
we use a \emph{per-user} pairwise ranking loss to generate a \emph{regressor} over $X$ specifying the
\emph{degree} to which items exhibit the
 soft attribute corresponding to tag $g$:
\begin{align}
\calL(\phi_g; (D_u)_{u\in\calU}) = \!\!\sum_u \sum_{\substack{i\in T_{u,g}\\ j\in T_{u,\olg}}} \!\!\log (1 + e^{- \phi_g^\top(\phi_I(j) - \phi_I(i))})\! +\! \frac{\lambda}{2} \phi_g^\top \phi_g.
\label{eq:rankingloss}
\end{align}
This logistic pairwise loss is the same as that used in RankNet \cite{RankNet2005}, hence we refer to this method as RankNet for CAVs.\footnote{Other loss functions---for instance, LambdaRank \cite{burges2010ranknet}, exponential loss, or loss hinge \cite{cao2006adapting}---may also be useful. We empirically examine LambdaRank in our experiments.}
The regressor $\phi_g$ obtained serves as our CAV. Notice that $\phi_g$ is linear in the learned item embedding features $\phi_I$.

Given a CAV $\phi_g$, the degree to which an item $i$ satisfies the induced attribute is given by the score $\phi_g(i) = \phi_g^\top \phi_I(i)$. The \emph{quality} $Q(\phi_g; D)$ of a CAV on data set $D$ is the fraction of the tag applications that it orders ``correctly'', i.e., if $i\!\in\! T_{u,g}, j\!\in\! T_{u,\olg}$, then $\phi_g(i)\! \geq\! \phi_g(j)$:
\begin{align}
Q(\phi_g;D) = \sum_u\! |\{(i,j)\! :\! i\!\in\! T_{u,g}, j\!\in\! T_{u,\olg}, \phi_g(i)\! \geq\! \phi_g(j)\}|\, / \, |D|.
\end{align}
This is closely related to the pairwise loss above and is a key metric we use to evaluate CAVs, and can be used as a measure of a CAV's ``usefulness'' for, say preference elicitation, as we discuss below.

\subsection{Nonlinear Attributes}
\label{sec:nonlinear}

A limitation of the linear approach is that if a CAV for tag $g$ is linear in the latent embedding space $X\subseteq \R^d$, every user's utility for $g$ is also linear in $X$. For example,
if the CAV for the tag `violent' is linear in $X$, it implies the preference for any user $u$ must be such that she prefers either \emph{maximally} or \emph{minimally} violent movies; she cannot prefer movies that are only ``somewhat violent'' and disprefer both movies that are more and less violent. For many soft attributes, this will not be the case---real-world preferences are often nonlinear (e.g., saturating \cite{french}) and even non-monotone (e.g., single-peaked \cite{moulin:PC80,regenwetter-BSC:2006}) with respect to many natural attributes. Such attributes are unlikely to be adequately represented linearly in embedding space $X$. Fortunately, CAVs can also be applied to nonlinear DNN representations.

We assume a two-tower/dual-encoder model and extract 
CAVs from hidden layers of the item DNN $N_I$.
For ease of exposition, we follow \citet{kimTCAV:icml18} by assuming that relevant concepts, if learned, can be uncovered within a single hidden layer of the (trained) deep collaborative filtering model.\footnote{\citet{kimTCAV:icml18} argue that TCAV is typically robust to this choice. We can apply TCAV to multiple layers simultaneously as well, a methods we we employ in some of our experiments below.} Given positive and negative examples as in Section~\ref{sec:linear}, we use activation $\phi_{I,\ell}(i)$ of the $\ell$'th layer of $N_I$ as item $i$'s training input representation instead of item embedding $\Phi_I(i)$.
Otherwise, the regressor is trained as above, and the validity of the induced CAV can be tested in the same fashion as well.

The result is a regressor $\phi_{g,\ell}$ that can be applied to an item's representation in the intermediate ``activation space'' $X^\ell_I$, where
$\phi_{g,\ell}^\top \phi_{I,\ell}(i)$ captures the degree to which item $i$ satisfies the attribute corresponding to tag $g$. The projection of this regressor through the last $L-\ell$ layers of
$N_I$ generates a (nonlinear) manifold in embedding space $X$. This gives considerably more flexibility to the relation between user utilities and soft attributes. In our nonlinear CAV experiments, we treat $\ell$ as a tunable hyper-parameter, and our reported results are based on the best CAVs extracted from these layers.

\subsection{Empirical Assessment of CAV Quality}
\label{sec:empiricalObjective}

We first evaluate our approach on synthetic data, which allows strict control over the generative process and direct access to ground truth labels (i.e., the degree of any attribute expressed by an item), as we explain below. We then test it on real-world data.
For linear soft attributes, we train a collaborative filtering model $\Phi = (\phi_U, \phi_I)$ using \emph{weighted alternating least squares (WALS)}
\cite{wals:icdm08},
with the following regularized objective:
\begin{align}
(\phi^*_U, \phi^*_I)\in\arg\min\,\,\sum_{u,i}c_{u,i}(\hatr_{u,i}-r_{u,i})^2\!+\!\kappa(||\phi_U||^2 \!+\!||\phi_I||^2).
\end{align}
Here $c_{u,i}$ is a confidence weight for the predicted rating $\hatr_{u,i}=\phi^\top_{U}(u;\theta_U) \phi_{I}(i;\theta_I)$, and $\kappa > 0$ is a regularization parameter. We select embeddings $(\phi^*_U, \phi^*_I)$ using validation loss and an item-oriented confidence weight
$c_{u,i}\propto n -\sum_u r_{u,i}$ (i.e., higher weight for less-frequently or lower-rated items) 
For nonlinear attributes, we train a two-tower DNN embedding model with SGD/Adam \cite{adam}.
Further details on synthetic data generation (\ref{app:syntheticdata}), the data set we use (\ref{app:movielens}), and training methods we adopt (\ref{app:training})
are provided in the appendix.

\vskip 2mm
\noindent
\textbf{Synthetic Data.}\hspace*{2mm}
We refer to Appendix~\ref{app:syntheticdata} for a detailed description of the generative model used to construct our synthetic data sets. At a high level,
to generate synthetic data a generative model outputs both user-item ratings $\bfR$ and tag data $\bfT$ for $n = 25,000$ users and $m = 10,000$ items. Users and items are each represented by $d=25$-dimensional embedding vectors,
sampled from pre-defined mixture distributions to induce correlation in the data. For linear utility, user ratings are generated by first sampling items (giving a sparse ratings matrix $\bfR$) and then their ratings (noise added to the user/item dot product). For nonlinear utility, utility is the sum of nonlinear sub-functions (one per dimension) peaked at some (random) point and dropping
as the item moves away from that peak.
In both cases, users 
are more likely
to rate items with higher utility.

To generate tags, five of the 25 latent item dimensions are treated as user-interpretable or ``taggable,'' each with a different tag.
Intuitively, these correspond to dimensions in the latent space that are both ``perceivable'' and ``interpretable'' by users.
Note, however, that the collaborative filtering model does not have access to the structure or details of the underlying generative process; it only sees the ratings and tag data so-generated.
Each user $u$ has a random \emph{propensity to tag}, which influences the probability with which they will tag an item they have rated.
Users are also more likely to tag higher-rated items---given their fixed tagging propensity, $u$ is more likely to tag a item they have rated highly. For a given tag $g$, there is a \emph{fixed} (non-subjective) threshold $\tau_g$ against which $u$ evaluates an item to be tagged, and noisily applies $g$ if that threshold is met.

\vskip 2mm
\noindent
\textbf{CAV Accuracy.} \hspace*{2mm}
We evaluate CAV accuracy---how well the learned regressors capture the underlying soft attribute---using standard metrics of prediction quality with respect to user tag usage on held-out test data. The synthetic model also allows evaluation relative to the \emph{ground-truth item representations and attribute levels of each tag} since we have access (for purposes of evaluation) to the \emph{true latent attribute values of each item} as generated by the model. Indeed, this is the main reason we include synthetic results since it provides a definite assessment of our techniques.
We evaluate three training methods, binary logistic regression, RankNet \cite{RankNet2005}, and LambdaRank \cite{burges2010ranknet}.
We split the tag data $\bfT$ into training and test sets and 
use the following metrics to measure the accuracy of CAVs on the test set:
\begin{enumerate}[(i)]
\item \emph{Accur}, the mean accuracy of the logistic model or \emph{quality} $Q(\phi_g; D)$ of the ranking model; 
\item \emph{Sprmn}, the \emph{Spearman rank correlation coefficient} between predicted and ground-truth attribute values.\footnote{We also computed precision and recall results, but these are omitted since they simply corroborate the findings generated by these metrics.}
\end{enumerate}
The latter compares the \emph{ranking} of all items by their \emph{ground-truth} attribute value with the ranking induced by the attribute's \emph{learned CAV}.
This metric applies equally to ``nonlinear'' latent attributes, since the CAV still provides a precise ordering of all items according to the attribute.

\vskip 2mm
\noindent
\textbf{Synthetic Results.} \hspace*{2mm}
Table~\ref{tab:non_subjective} shows the predictive performance of our CAVs on synthetic data for three 
different combinations of user data model and general CAV approach: (i) user utility is linear and we train linear CAVs; (ii) user utility is nonlinear but we train linear CAVs (Lin-Emb); and (iii) user utility is nonlinear and we train nonlinear CAVs (NL-Emb).
We list the accuracy metrics above, where CAVs are trained using either logistic regression, RankNet or LambdaRank, 
with results averaged over the five tags. We bold the best value within each of the three settings.
We see that the CAVs predict user tagging behaviors reasonably accurately as indicated by \emph{Accur}, and reliably order test items according to the ground-truth value of the underlying soft attribute, as evidenced by \emph{Sprmn}, despite the noise in the tagging process.
The ranking methods, RankNet and LambdaRank, dominate logistic regression with respect to both \emph{Accur} and \emph{Sprmn}, which suggests that accounting for variation in user tagging behavior is important (we discuss this further when describing subjective results in the next section).
For nonlinear utilities, we also compare the best ``linear'' CAV (extractable from the output embedding) with the best nonlinear CAV (extractable from DNN hidden layers). While linear CAVs may be viewed as conceptually simpler, they are mismatched to the underlying attributes when user utility is nonlinear. Unsurprisingly, nonlinear CAVs outperform their linear counterparts, demonstrating the value of seeking nonlinear (or ``distributed'') attribute representations within the DNN, and the power of TCAV to help interpret them. We also include a baseline tag recommender \emph{PITF (pairwise interaction tensor factorization)} \cite{rendle2010}, which uses tensor decomposition to model pairwise interactions between users, items, and tags (see more details in the appendix)---note that it produces linear representations. Its tag prediction accuracy is worse than that of the CAV approaches. \emph{PITF} does not use ratings which may be valuable in training personalized representations with sparse tag data. 

\begin{table}[t]
\begin{minipage}{0.55\textwidth}
  \centering
  {\footnotesize
  \begin{tabular}{|c||c|c|c|c|c|c|}
  \hline
Utility     &\multicolumn{2}{c|}{\textbf{Linear}} & \multicolumn{2}{c|}{\textbf{Linear}} & \multicolumn{2}{c|}{\textbf{NonLinear}}\\
CAV Model     & \multicolumn{2}{c|}{\textbf{Lin-Emb}} & \multicolumn{2}{c|}{\textbf{Lin-Emb}} & \multicolumn{2}{c|}{\textbf{NL-Emb}}\\
     & \, Accur\, & \, Sprmn\, & \, Accur\, & \, Sprmn\, & \, Accur\, & \, Sprmn\, \\ \hline\hline
 Log.\ Regr. & 0.906  &	0.569 &  0.889 & 0.565	 &  0.922 & 0.577	\\ \hline
 RankNet & \textbf{0.968} &	0.674 & 0.943 & \textbf{0.670}	 &  \textbf{0.978} & \textbf{0.686} 	 \\ \hline
 LambdaNet & 0.961 &	\textbf{0.679} & \textbf{0.947} & 0.666	 &  0.974 & 0.680 \\ \hline	
 PITF & 0.683 & 0.056 & 0.707 & 0.070 & N/A & N/A \\ \hline
 \end{tabular}
 \vspace*{2mm}
  \caption{CAV Evaluation, Synthetic Data (Non-subjective)}
  \label{tab:non_subjective}
  }
   \vspace{-0.2in}
\end{minipage}
\hfill
\begin{minipage}{0.4\textwidth}
  \centering
  {\footnotesize
  \begin{tabular}{|c||c|c|}
  \hline
CAV Model     & {\textbf{Lin-Emb}} & {\textbf{ NL-Emb}}\\
     & \, Accur\, & \, Accur\,\\ \hline\hline
 Log.\ Regr. &  0.727  & 0.745	\\ \hline
 RankNet &  0.803 &	\textbf{0.820} \\ \hline
 LambdaNet & \textbf{0.804} & 0.818	\\ \hline
 PITF & 0.715 & N/A \\ \hline
 \end{tabular}
  }
  \vspace*{2mm}
    \caption{CAV Evaluation, MovieLens}
   \vspace{-0.2in}
  \label{tab:non_subjective_movielens}
  \end{minipage}
\end{table}

\vskip 2mm
\noindent
\textbf{MovieLens Results.} \hspace*{2mm}
We also evaluate our methods on the more realistic MovieLens20M data set \cite{harper16:movielens}. In contrast to the synthetic results---which are most useful for evaluating the \emph{effectiveness} of our methods in controlled settings---evaluation on real-world data sets like MovieLens20M, and the rater-based evaluation in Section~\ref{sec:user_study}, are more suitable for testing their \emph{applicability} ``in the wild.''
MovieLens20M comprises 
20M movie ratings (1--5 scale with half-star increments) and 465K tag-instances  applied to 27K movies by 138K users.
Tags are user-generated strings that describe various types of movie attributes. These can include attributes such as genres like `sci-fi', `comedy', `action';  descriptions like `emotional', `funny', `atmospheric'; and themes like `zombies', `world war 2', `cyberpunk'.
We split rating and tag data into train and test sets such that \emph{all examples} for any specific user-item pair are present in exactly one of these subsets (i.e., if $u$ rated and tagged $i$, the rating and \emph{all tags $g$ applied by $u$ to $i$} are all in either the training set or all in the test set). We generate $d=50$-dimensional user and item embeddings: we use 
WALS as our collaborative-filtering method in the case of linear CAVs, and train two-tower DNNs for non-linear CAVs.

Positive examples for
tag $g$ are user-item pairs to which $g$ has been applied; negatives are those tagged by that user, but not with $g$ (recall that we use negative examples of $g$ for user $u$ \emph{only if} $u$ has positively applied $g$ to at least one item). 
Given the spareseness of tags---of 30745 unique tags, more than 28000 are applied to fewer than 10 unique movies---to ensure that the analyzed tags are relevant to sufficiently many user-item interactions, 
we train CAVs only for those tags that are among the 250 most frequently applied both by unique users and to unique items, which results in 164 tags.
Table~\ref{tab:non_subjective_movielens} shows test accuracy for linear and nonlinear CAVs. As in the synthetic settings, the ranking methods outperform logistic regression, which again hints at some subjectivity (see next section).
While we cannot measure Spearman correlation (since we have no ground truth ranking), nor control the form of user utility, we see that nonlinear CAVs perform slightly better than linear CAVs, suggesting that user preferences for at least some MovieLens tags are nonlinear in their embedding-space representation.
Again, as in the synthetic settings, our CAV methods consistently outperform PITF.

\section{Identifying Subjective Attributes}
\label{sec:subjectiveCAVs}

If users largely agree on the usage of tags, it is reasonable to treat the semantics of a tag as a \emph{single} soft attribute or CAV as we do above.
But in many cases, different users may have different ``senses'' in mind when they apply a tag. For example, one user may use the term `funny' to describe  movies that are silly, involving, say, physical or slapstick humor, while another may use the same term to refer to dry, political satire.\footnote{Some users may offer more ``refined'' tags to distinguish their intended sense, but many will not. For this reason, we still require the ability to disambiguate a user's intended meaning in the general case.}
While correlated, these two \emph{tag senses} will order movies quite differently. Such \emph{sense subjectivity} may hinder our ability to produce an accurate CAV and understand a user's true intent. We now turn to this issue and show how to learn \emph{personalized semantics} for tags.

\subsection{Subjectivity of Degree}
\label{sec:subjectiveDegree}

As discussed above, \emph{degree subjectivity} is likely to emerge quite naturally. The use of \emph{intra-user pairwise comparisons} with a ranking loss in CAV training (see Eq.~(\ref{eq:rankingloss})) ensures that the induced CAVs are robust to this form of subjectivity. However, since two users may use a tag $g$ differently if they have different thresholds for applying $g$, interpreting user $u$'s usage (e.g.\ statements such as ``that's not funny'' or ``I'm in the mood for something funny'') requires a \emph{personalized semantics} that is sensitive to her threshold. 
Let $g$ be a tag that is degree subjective, $\phi_g$ be $g$'s CAV and $\phi_g(i)$ be the degree to which item $i$ satisfies $\phi_g$. A \emph{user-specific threshold} $\tau^u_g \in \R$ determines a user-specific semantics for $g$: tag $g$ applies (typically, noisily) to item $i$ for user $u$ only if $\phi_g(i) \geq \tau^u_g$. Equivalently, this can be viewed as a \emph{personal} linear separator for $u$ in $X$, but constrained to be orthogonal to the direction $\phi_g$ induced from the \emph{population} labels. The (estimated) \emph{optimal personal threshold} $\tau^u_g$ minimizes the number of misclassifications with respect to $u$'s usage of $g$:
\begin{align}
\tau^u_g \in \arg\min_\tau |\{i\in T_{u,g} : \phi_g(i) \geq \tau\} \cup \{i\in T_{u,\olg} : \phi_g(i) < \tau\}|.
\end{align}

Among the continuum of such minimizers, the threshold $\tau^u_{g}$ that maximizes the margin between the nearest bracketing positive and negative items is a natural choice for $u$'s personal semantic threshold. Of course, since tag usage by an individual user is typically extremely sparse, these thresholds are likely to be noisy---this is one reason that we do not compute CAVs themselves are on a per-user basis.
However, we can refine $u$'s semantics with well-chosen queries, e.g., ``Do you consider movie $m$ to be violent?'' This can be used to implement a loose binary search for an approximate threshold (possibly made robust to account for noisy responses), but we defer this to future research. If usage is correlated within user sub-populations, generalization of thresholds across users is also viable, as we discuss in the case of sense subjectivity below.

\subsection{Subjectivity of Sense}
\label{sec:subjectiveSense}

We now turn to \emph{sense subjectivity}. We can readily detect sense subjectivity and assign a \emph{personalized semantics} for a tag $g$ to different (groups of) users, using \emph{distinct CAVs} for each tag sense. We assume that $g$ has \emph{at most} $s_g$ distinct senses $g[1],\ldots, g[s_g]$, for some small 
positive integer $s_g$, where each sense denotes a different soft attribute (we discuss their relation below). Moreover, each user adopts exactly \emph{one} such sense of $g$.\footnote{We defer to future work the possibility of \emph{mixtures} of senses.}
We propose a method to discover whether a tag has multiple senses, and to uncover suitable CAVs for each sense if so.

Intuitively, if the CAV quality metric $Q(\phi_g; D)$ is high, then CAV $\phi_g$ does a good job of explaining usage of tag $g$ among all users in data set $D$. If not, then model $\Phi$ is unlikely to have learned a good representation for $g$. This could be due to $g$ being poorly correlated with user ratings (hence, user preferences), or because $g$ has multiple senses. In the latter case, if $g$ has $s$ senses, there should be some \emph{user-partitioning} of $D$ into subsets $D_1, \ldots, D_s$ of users such that there is a CAV $\phi_{g,k}$ with high quality $Q(\phi_{g,k}; D_k)$ for each $k\leq s$. We first propose a simple scheme to find a good set of CAVs for a fixed $s$, then discuss determination of a suitable number of senses $s \leq s_g$.

Assume a fixed number of target senses $s$ and a given data set $D$. We propose a simple scheme for generating $s$ CAVs corresponding to maximally distinct senses for tag $g$. Let $\Sigma = \{\sigma_1, \ldots, \sigma_s\}$ be a partitioning of users into $s$ clusters, where $\sigma_k$ is the set of users that, we assume for the time being, adopt a common sense for $g$. Let $D_k$ be the restriction of $D$ to tag data generated by users $u\in\sigma_k$. For a fixed $\Sigma$, we can readily generate a CAV $\phi_{g,k}$ for each data set $D_k$ capturing the corresponding sense, and measure its quality $Q(\phi_{g,k};D_k)$. Of course, this quality depends on whether the partitioning $\Sigma$ is sensible (i.e., whether most users in each cluster use $g$ similarly). If the quality of these CAVs is low, one can repartition users by ``assigning'' each user $u$ to the cluster in $\Sigma$ whose CAV best explains her tag usage:
\begin{align}
k^*_u = \arg\max_k |\{(i,j)\! :\! i\!\in\! T_{u,g},\; j\!\in\! T_{u,\olg},\; \phi_{g,k}(i)\! \geq\! \phi_{g,k}(j)\}|.
\label{eq:bestcluster}
\end{align}

This leads to an EM-like alternating optimization procedure \cite{bishop_PRML:2006} for finding a good clustering that repeatedly applies the following steps:
\begin{enumerate}[(a)]
\item Learns a CAV $\phi_{g,k}$ for tag $g$ for each current (user) cluster $k$ using one of our methods above;
\item Reconstructs the clusters by assigning each user to the cluster whose new CAV $\phi_{g,k}$ best explains her usage of tag $g$ per Eq.~\ref{eq:bestcluster}; i.e., $D_k \leftarrow \{u \in\calU : k^*_u = k\}$.
\end{enumerate}
The iterative process proceeds until the partitioning $\Sigma$ no longer changes or quality improvements become sufficiently small.
Since we use a hard clustering scheme and only change $\Sigma$ when the new clusters yield strictly better quality, it is easy to see that the EM procedure terminates in a finite number of steps. If we assign each $u$ by minimizing the logistic or ranking loss incurred by $u$, using convergence properties of the standard $k$-means algorithm (e.g.,~\cite{bottou_kmeans}), one can show that the procedure converges to a local minimum and generates $s$ distinct CAV senses.\footnote{A number of variants of this general approach can be explored (but are beyond the scope of this work): various criteria for initialization (e.g., using ICA for initial clustering); hard vs.\ soft clustering; various termination criteria; etc.}

\begin{table}[t]
  \centering
  {\footnotesize
  \begin{tabular}{|c||c|c|c|c|c|c|}
  \hline
Utility     & \multicolumn{2}{c|}{\textbf{Linear}} & \multicolumn{2}{c|}{\textbf{NonLinear}} & \multicolumn{2}{c|}{\textbf{NonLinear}}\\     
CAV Model & \multicolumn{2}{c|}{\textbf{Lin-Emb}} & \multicolumn{2}{c|}{\textbf{Linear-Emb}} & \multicolumn{2}{c|}{\textbf{NL-Emb}}\\
     & \, Accur.\, & \, Sprmn\, & \, Accur.\, & \, Sprmn\, & \, Accur.\, & \, Sprmn\, \\ \hline\hline
 Log.\ Regr. & 0.872  &	0.566 &  0.860 & 0.523	 &   0.886 & 0.548	\\ \hline
 RankNet & 0.960 & \textbf{0.671}	 &  \textbf{0.947} & \textbf{0.660}	 &  \textbf{0.961} & 0.680 \\ \hline
 LambdaNet & \textbf{0.962} & 0.669	 &  0.938 & 0.653	 &  0.958 & \textbf{0.684} \\ \hline
 PITF & 0.700 & 0.064 & 0.708 & 0.068 & N/A & N/A \\ \hline
 \end{tabular}
  } \\
  \vspace*{2mm}
  \caption{CAV Evaluation, Synthetic Data (Degree subjectivity)}
  \label{tab:subjectivity_degree}
     \vspace{-0.2in}
\end{table}

Assuming that the complexity of the regression/ranking sub-problem is linear in number of examples $N$, with $L$ iterations of EM, the complexity of this approach is $O(LNs)$.
Empirically the EM method converges very quickly ($L \leq 5$) in 
both our MovieLens and the synthetic data experiments below, so the computational cost of the EM algorithm is minimal (and recall that these are offline computations).

We can search for the appropriate number of senses---effectively implementing a form of \emph{model selection} \cite{bishop_PRML:2006}---by starting with an initial (single) CAV, and applying the procedure above to gradually increasing number of clusters $s = 2, 3, \ldots, s_g$, and terminating once the improvement in average quality, $\sum_k (|D_k|/|D|) Q(\phi_{g,k}; D_k)$ is negligible.\footnote{A number of other more complex approaches can be considered.}

Notice that our approach to disentangling tag senses is in the style of top-down ``disaggregative'' or \emph{divisive}  \cite{manning:introductionIR2008}. We do this since bottom-up agglomerative clustering \cite{manning:introductionIR2008} is likely to be very noisy---the tag set of any individual user is extremely sparse, so attempts to produce a CAV for very small groups of users will generally be unreliable. 
Once we find the appropriate number of senses and corresponding CAVs, in practice, we can continue to update them as new users, items and tagging data arises. For example, given a new user $u$ who has applied tags to items, we can associate a ``personal'' semantics for $u$ with a given tag $g$ by assigning $u$ to the $g$-cluster whose CAV best fits $u$'s usage of data by computing $k^*_u$.

\subsection{Empirical Assessment}
\label{sec:empiricalSubjective}

As above, we test our methods for subjective soft attribute identification on both synthetically generated data with ground truth labels, and on the MovieLens20M data set.

\begin{table}[t]
  \centering
  {\footnotesize
  \begin{tabular}{|c||c|c|c|c|c|c||c|c|c|c|c|c|}
    \hline
     & \multicolumn{6}{c||}{\textbf{\normalsize Subjective Tags}} & \multicolumn{6}{c|}{\textbf{\normalsize Objective Tags}} \\
     & \multicolumn{6}{c||}{\, } & \multicolumn{6}{c|}{\, } \\
Utility     & \multicolumn{2}{c|}{\textbf{Linear}} & \multicolumn{2}{c|}{\textbf{NonLinear}} & \multicolumn{2}{c||}{\textbf{NonLinear}} & \multicolumn{2}{c|}{\textbf{Linear}} & \multicolumn{2}{c|}{\textbf{NonLinear}} & \multicolumn{2}{c|}{\textbf{NonLinear}}\\
CAV Model          & \multicolumn{2}{c|}{\textbf{Lin-Emb}} & \multicolumn{2}{c|}{\textbf{Lin-Emb}} & \multicolumn{2}{c||}{\textbf{NL-Emb}} & \multicolumn{2}{c|}{\textbf{Lin-Emb}} & \multicolumn{2}{c|}{\textbf{Lin-Emb}} & \multicolumn{2}{c|}{\textbf{NL-Emb}}\\
     &  Accur. &  Sprmn &  Accur. &  Sprmn &  Accur. &  Sprmn &  Accur. &  Sprmn &  Accur. &  Sprmn &  Accur. &  Sprmn \\ \hline\hline
 Log.\ Regr. & 0.643 & 0.039 & 0.634 & 0.026 & 0.616 & 0.036 & 0.936 & 0.634 & 0.926 & 0.642 & 0.922 & 0.635 \\ \hline
 EM Log.\ Regr. & 0.833 & \textbf{0.478} & 0.796 & 0.419 & 0.828 & 0.476 & 0.937 & 0.631 & 0.926 & \textbf{0.637} & 0.922 & 0.635 \\ \hline
 RankNet & 0.615 & 0.042 & 0.606 & 0.035 & 0.603 & 0.022 & \textbf{0.992} & \textbf{0.636} & \textbf{0.987} & 0.636 & \textbf{0.982} & \textbf{0.636} \\ \hline
 EM RankNet & \textbf{0.922} & 0.466 & \textbf{0.915} & \textbf{0.468} & \textbf{0.908} & \textbf{0.468} & \textbf{0.992} & 0.633 & \textbf{0.987} & 0.633 & \textbf{0.982} & \textbf{0.636} \\ \hline
  LambdaRank & 0.615 & 0.028 & 0.606 & 0.036 & 0.604 & 0.009 & \textbf{0.992} & 0.634 & \textbf{0.987} & 0.632 & \textbf{0.982} & 0.634 \\ \hline
 EM LambdaRank & 0.920 & 0.458 & \textbf{0.915} & 0.465 & \textbf{0.908} & 0.465 & \textbf{0.992} & 0.631 & \textbf{0.987} & 0.630 & \textbf{0.982} & 0.631 \\ \hline
 PITF & 0.672 & 0.063 & 0.711 & 0.070 & N/A & N/A & 0.640 & 0.050 & 0.675 & 0.074 & N/A & N/A \\ \hline
 \end{tabular}
  } \\
  \vspace*{2mm}
   \caption{CAV Evaluation, Synthetic Data (Sense Subjectivity): Subjective Tags (left half), Objective Tags (right half).}
  \label{tab:subjectivity_sense}
     \vspace{-0.2in}
\end{table}

\vskip 2mm
\noindent
\textbf{Synthetic Results.} \hspace*{2mm}
We again test our approach on synthetic data to exploit access to ground truth CAV semantics for our tags for purposes on evaluation.
We generate several data sets incorporating both degree and sense subjectivity, as well as linear and nonlinear soft attributes.
We refer to App.~\ref{app:syntheticdata} for a detailed description of the generative model used to construct these data sets with $n = 25,000$, $m = 10,000$, and $d=25$.
The model is similar to that used in Section~\ref{sec:empiricalObjective}, differing only in the addition of subjectivity to the user tagging behavior.

To test degree subjectivity, we use five tags as above, but with each user's personal tagging threshold sampled from a mixture distribution with two components.
To test sense subjectivity, we introduce a subjective tag ``tag-S'' with three senses,
each reflecting one of three (of the five) taggable dimensions. As above, each of these is associated with a specific semantics in embedding space. Each user $u$ adopts exactly one of these three senses; specifically, when applying tag-S to any item, they assess that it based on the specific sense of that tag (of the three) to which they have been assigned. The remaining two tags are objective. As above, ratings and tags are applied in a noisy fashion to generate the synthetic data set.

We evaluate the three CAV training methods used in Section~\ref{sec:empiricalObjective}, applying each to a linear (WALS) model and a nonlinear two-tower model as needed.
For sense subjectivity, we test our EM-like algorithm with each training method. Further details on data generation and the experimental set up can be found in the appendix.

Table~\ref{tab:subjectivity_degree} summarizes performance of the CAVs under degree subjectivity, using the same methods and models as in Section~\ref{sec:empiricalObjective} (results averaged over the five degree-subjective tags).
In contrast to the non-subjective case in Section~\ref{sec:empiricalObjective}, where users have the same threshold for each tag, here the per-user ranking-based methods (RankNet and LambdaRank) significantly outperform logistic regression, demonstrating the need to be sensitive to a user's degree subjectivity.

Table~\ref{tab:subjectivity_sense} summarizes  results for sense subjectivity, showing CAV accuracy for our baseline methods both with and without our EM-based approach for distinguishing senses. The left side of the table shows results for the sense-subjective tag-S, demonstrating that our EM procedure can dramatically improve CAV accuracy by reliably disentangling the three distinct senses of tag-S. This demonstrates that treating a subjective concept as if it were objective can be problematic, leading to erroneous interpretations of any given user's intent when using that tag. Note also that our ranking methods perform better than logistic regression
(\emph{EM-RankNet} and \emph{EM-LambdaRank} perform similarly).
The right side of the table shows CAV accuracy on the two \emph{objective} tags. We see that the EM and non-EM methods perform almost identically. This is important since it suggests our techniques are unlikely to identify spurious subjective senses of a tag if in fact tag usage is objective.
Somewhat intriguingly, we see that the use of nonlinear CAVs does not offer much improvement over linear CAVs with ranking methods, though nonlinear CAVs perform better when trained using logistic regression. The performance of the PITF baseline is worse than that of the CAV approaches in both the degree and sense subjectivity experiments.

\vskip 2mm
\noindent
\textbf{MovieLens Results.} \hspace*{2mm}
We also evaluate our subjective CAV methods on MovieLens20M. To assess degree subjectivity, we select 13 tags that we expect to exhibit various degrees of subjectivity and compare the accuracy of different CAV methods for each (the tags are listed in the top row of Table~\ref{tab:movielens_subj13}). Since MovieLens data has no ground truth with respect to possible subjectivity,
Spearman rank correlation cannot be measured, hence we focus on CAV predictive accuracy only. 
Table~\ref{tab:movielens_subj13} shows Accuracy measures for the linear and nonlinear variants of our three primary methods for the 13 MovieLens tags.
Generally, our ranking methods outperform logistic regression, suggesting that real users exhibit some variation in their thresholds (degree subjectivity). We also see that some tags (e.g., \emph{sci-fi}) have much higher agreement and CAV-predictability across users than others (e.g., \emph{funny}), hence arguably less degree subjectivity.  In particular, of the 13 tags, across all models, \emph{funny}, \emph{surreal}, \emph{quirky} and \emph{atmospheric} seem to be most subjective, while \emph{sci-fi}, \emph{action} and \emph{classic} are the most objective.

We also observe differences in the degree of improvement offered by nonlinear CAVs vs.\ linear CAVs across the tags: those with larger improvements (e.g., \emph{dark comedy}, \emph{dystopia}) suggest that user utility may often be nonlinear in the degree of that attribute (so extreme degrees may not be preferred); while those where nonlinear CAVs perform no better, or even worse (e.g., \emph{sci-fi}, \emph{action}, \emph{funny}), may be most preferred at their maximum or minimum degree.

\begin{table}[t]
    \centering
    \scalebox{0.70}{
    \begin{tabular}{|c||c|c|c|c|c|c|c|c|c|c|c|c|c|}
    \hline
    &sci-fi & \begin{tabular}{@{}c@{}}atmo- \\ spheric\end{tabular} & surreal & \begin{tabular}{@{}c@{}}twist \\ ending \end{tabular} & action & funny & classic & \begin{tabular}{@{}c@{}}dark \\ comedy\end{tabular} & quirky & \begin{tabular}{@{}c@{}}psych- \\ ology \end{tabular} & dystopia & stylized & \begin{tabular}{@{}c@{}}thought- \\ provoking\end{tabular} \\ \hline\hline
      Log.\ Regr.   & 0.831 & 0.721 & 0.739 & 0.705 & 0.823 & 0.689 & 0.818 & 0.714 & 0.725 & 0.715 & 0.737 & 0.753 & 0.764  \\ \hline
      RankNet   & 0.905 & 0.793 & 0.811 & 0.817 & \textbf{0.899} & \textbf{0.775} & 0.877 & 0.822 & \textbf{0.840} & 0.812 & 0.850 & 0.866 & 0.834 \\ \hline
      LambdaNet & \textbf{0.906} & 0.784 & 0.788 & 0.869 & 0.876 & 0.605 & 0.838 & 0.830 & 0.821 & 0.788 & 0.867 & 0.809 & 0.839  \\ \hline
     NL. Log.\ Regr.   & 0.831 & 0.725 & 0.742 & 0.711 & 0.812 & 0.681 & 0.821 & 0.705 & 0.751 & 0.704 & 0.807 & 0.811 & 0.764  \\ \hline
     NL. RankNet   & 0.893 & 0.838 & \textbf{0.827} & 0.854 & 0.890 &  0.754 & \textbf{0.888} & 0.868 & 0.833 & \textbf{0.837} & 0.882 & 0.856 & 0.844 \\ \hline
     NL. LambdaNet & 0.891 & \textbf{0.843} & 0.807 & \textbf{0.875} & 0.880 & 0.744 & 0.865 & \textbf{0.891} & 0.825 & 0.796 & \textbf{0.921} & \textbf{0.898} & \textbf{0.856} \\ \hline
    \end{tabular}
    }\\
    \vspace*{2mm}
   \caption{CAV Accuracy Evaluation, 13 Possible Subjective Concepts in MovieLens}
   \label{tab:movielens_subj13}
\vspace{-0.2in}
\end{table}

We have no labeled data in MovieLens20M with which to assess sense subjectivity. So to evaluate this form of
subjectivity, we construct two types of \emph{artificial tags} from MovieLens data.
First are \emph{objective tags}, capturing four \emph{genres} (comedy, horror, fantasy and romance). For a random subset of user-item pairs in the tag data set, we add the corresponding user-item-genre triple to the set if the item's meta-data lists that genre. This ensures that the new ``genre tag'' data replicates natural tagging patterns. We also add a synthetic tag \textit{odd year}---was a movie was released in an even or odd year---to 50\% of user-item pairs. This (presumably) \emph{preference-irrelevant} attribute acts as a baseline for which no good CAV should be discoverable. These tags are ``objective''---their presence does not depend on a user's interpretation of tags (though they may depend on a user's inclination to tag certain types of movies).

The second artificial tag type are \emph{sense-conflated tags}, constructed by coalescing several related ``ground'' tags into a single ``meta-tag'' then replacing each ground tag with that meta-tag. Each ground tag in the group can be viewed as a subjective sense of the meta-tag.
We test our ability to ``disentangle'' the different senses of the meta-tag relative to the ground truth. We introduce four meta-tags:
\begin{itemize}
    \item Meta-tag \textit{monsters}: groups ground tags \textit{zombies}, \textit{ghosts} and \textit{vampires};
    \item Meta-tag \textit{funny}: groups ground tags \textit{parody}, \textit{satire}, and \textit{dark humor};
    \item Meta-tag \textit{intrigue}: groups ground tags \textit{corruption}, \textit{conspiracy}, and \textit{politics}; and
    \item Meta-tag \textit{relationship}: groups ground tags \textit{family}, \textit{friendship}, and \textit{love story}. 
\end{itemize}
For each user and meta-tag, we choose \emph{exactly one} ground tag from the group as that user's \emph{designated sense} and add the meta-tag to the data set for each user-item-tag triple that uses the ground tag. For example, some users have all of their \textit{friendship} tags replaced with the meta-tag \textit{relationship}, while others have all of their \textit{love story} tags replaced this way.
In this way, we have access to subjective tags that follow the natural tag application patterns in the MovieLens data, but with a ground-truth partitioning of users and (partial) item labeling according to the different senses of the meta-tag.

Table~\ref{tab:subjectivity_movielens} summarizes the accuracy of our trained CAVs for these two types of artificial tags. For the four ``objective'' \textit{genre} tags, the ranking-based methods outperform logistic regression. However, adding EM to our ranking methods provides only modest incremental benefit (especially relative to lift it offers for the ``subjective'' sense-conflated tags below). This implies that \textit{genres} exhibit at most modest, if any, subjectivity of sense, as expected.\footnote{EM will often provide some improvement in accuracy, even if distinct senses do not exist, by allowing some overfitting. In this case, EM may also overcome the limitations of the linear CAVs if user utilities for these objective genres are non-linear.} The `horror' and `fantasy' tags are the easiest to learn, suggesting they are more ``objective'' and ``linear.''

None of our methods uncovers a good CAV for the artificial tag \textit{odd year}---their predictive accuracy is barely above random. This is an important finding, since it corroborates our hypothesis that CAVs are useful for identifying \emph{preference-related} attributes/tags.
On the other hand, results on our four \emph{sense-conflated} tags clearly demonstrate the ability of our EM-style approach to disentangle the distinct subjective or personal senses of each of the meta-tags---this is true for each baseline algorithm---thus greatly improving tag prediction accuracy.

\begin{table}[t]
  \centering
  {\footnotesize
  \begin{tabular}{|@{\ }c||@{\ }c|@{\ }c|@{\ }c|@{\ }c|@{\ }c||@{\ }c|@{\ }c|@{\ }c|@{\ }c|} \hline
 & \, Odd Year \, & \, Comedy\, & \, Horror\, & \, Fantasy\, & \, Romance \, & \, Monsters \, & \, Funny  \, & \, Intrigue \, & \, Relationship\, \\ \hline\hline
Log.\ Regr., Lin.\ Emb. & 0.519 & 0.521 & 0.770 & 0.685 & 0.693 & 0.671 & 0.658 & 0.669 & 0.662 \\ \hline
EM Log.\ Regr., Lin.\ Emb. & 0.532 & \textbf{0.685} & 0.759 & 0.744 & 0.704 & 0.831 & 0.769 & 0.712 & 0.730 \\ \hline
RankNet, Lin.\ Emb. & 0.505 & 0.620 & 0.790 & 0.778 & 0.730 & 0.718 & 0.705 & 0.660 & 0.634 \\ \hline
EM RankNet, Lin.\ Emb. & \textbf{0.593} & 0.676 & 0.833 & \textbf{0.824} & 0.749 & \textbf{0.892} & \textbf{0.874} & 0.834 & 0.840 \\ \hline
LambdaRank, Lin.\ Emb. & 0.533 & 0.609 & 0.809 & 0.779 & 0.716 & 0.719 & 0.718 & 0.661 & 0.623 \\ \hline
EM LambdaRank, Lin.\ Emb. & 0.582 & 0.670 & \textbf{0.838} & 0.819 & \textbf{0.762} & 0.883 & 0.870 & \textbf{0.836} & \textbf{0.847} \\ \hline
 \end{tabular}
  }\\
\vspace*{2mm}
\caption{CAV Accuracy Evaluation, Artificial MovieLens Tags (5 objective, 4 sense-conflated)}
  \label{tab:subjectivity_movielens}
    \vspace{-0.2in}
\end{table}

\section{CAV Evaluation with Rater Data}
\label{sec:user_study}

Our evaluation of CAV semantics in the preceding sections used two distinct approaches. Evaluation using synthetic data allows assessment relative to fully known, ground-truth labels (i.e., the degree of a soft attribute exhibited by each item is known), but does not capture directly how real users of recommender systems use or interpret the terms or tags associated with these attributes.
Our second approach, using MovieLens20M tagging behavior, does reflect actual user tag usage, but requires that we make some assumptions (e.g., negative tag imputation) about tag usage to draw specific inferences about the semantics of the underlying soft attributes. While both allow meaningful conclusions to be reached, neither reflect the ideal evaluation in which users explicitly tell us whether they believe one item exhibits more or less of a soft attribute than another.

In this section, we evaluate CAVs using precisely such rater data. Specifically, we use the \emph{SoftAttributes} data set~\cite{sigir21:filipandkristian},\footnote{See \texttt{https://github.com/google-research-datasets/soft-attributes}} in which raters explicitly compare movies based on the degree to which they exhibit specific attributes (we outline the precise nature of this data set below). This provides explicit pairwise comparisons by human raters against which to evaluate the CAVs learned from tagging or comparison data, and does not require any assumptions about usage, or any form of imputation or down-sampling. We first describe the \emph{SoftAttributes} data set in Section~\ref{sec:ratereval_setup}. In Section~\ref{sec:ratereval_MovieLens}, we use this rater data to evaluate the predictive accuracy of the CAVs we trained (see above) using MovieLens tags to provide an independent assessment of those CAVs that does not rely on the assumptions used to train the model. In Section~\ref{sec:ratereval_SA}, we train a new set of CAVs directly on the \emph{SoftAttributes} data set itself to determine if CAVs trained using \emph{explicit} (rather than imputed) soft-attribute comparisons have greater predictive accuracy. This latter evaluation also demonstrates the application of CAVs to tagging/attribute data generated in a different fashion (since the \emph{SoftAttributes} tag vocabulary is extracted from user reviews rather than explicit tagging of items).

\subsection{The \emph{SoftAttributes} Data Set}
\label{sec:ratereval_setup}

\citet{sigir21:filipandkristian} provide one of the first detailed investigations of the semantics of soft attributes in recommender systems, with an emphasis on the weaknesses of typical binary tagging-based approaches, including the fact that items often exhibit attributes to varying degrees, as well as the contextual, subjective nature of attribute usage. Apart from algorithmic contributions (which we use as a benchmark below), \citet{sigir21:filipandkristian} also created the \emph{SoftAttributes} data set.

The data set itself is generated using an intuitive process designed to extract robust (and personal) soft-attribute comparisons from raters. A rater $r$ is first asked to indicate a subset of ``familiar'' movies (i.e., movies they have seen) from a pool of popular movies (namely, the 300 most popular movies in MovieLens-20M). The rater $r$ is then presented with a specific soft attribute/tag $g$ (e.g., `violent'), an anchor movie $m_{r,g}$ and a set of 10 candidate movies $C_{r,g} = \{c_{i,r,g} : i\leq 10\}$, where the anchor and candidates are all drawn from $r$'s familiar set. Rater $r$ is then asked, using an intuitive graphical interface, to specify, for each candidate $c_{i,r,g}$, whether it is more, less, or about the same as the anchor $m_{r,g}$ w.r.t.\ $g$ (e.g., whether $c_{i,r,g}$ is more violent than $m_{r,g}$, less violent than $m_{r,g}$ or about the same as $m_{r,g}$ in terms of degree of violence). This divides the 11 movies $C_{r,g} \cup \{m_{r,g}\}$ into three equivalence classes: those movies $L_{r,g}$ that are ``less $g$'' than the anchor $m_{r,g}$, those $M_{r,g}$ that are ``more $g$,'' and those $S_{r,g}$ that are about the same. This is repeated for each of 60 soft attributes across 100 raters.\footnote{We refer to \citet{sigir21:filipandkristian} for details on how their soft attributes are generated and selected. Of note is the fact that they are extracted from movie reviews in an Amazon reviews corpus rather than from explicit tagging of items.} The resulting data set consists of approximately 250K pairwise comparisons of movies over these 60 soft attributes.

We use the data set in a similar fashion to \citet{sigir21:filipandkristian} with three classes of comparisons. For any fixed rater $r$ and tag $g$, we say:
\begin{itemize}
    \item $c \succ^s_{r,g} c'$ if $c\in M_{r,g}$ and $c' \in L_{r,g}$. These are \emph{strong differences}, since $c$ is more $g$ than $m_{r,g}$ and $c'$ is less $g$, so $c$ is ``two levels'' more $g$ than $c'$.
    \item $c \succ^w_{r,g} c'$ if $c\in M_{r,g}$ and $c' \in S_{r,g}$; or if $c\in S_{r,g}$ and $c' \in L_{r,g}$ These are \emph{weak differences}, since $c$ is only ``one level'' more $g$ than $c'$. 
    \item $c \approx^s_{r,g} c'$ if $c,c'\in M_{r,g}$ or $c,c' \in S_{r,g}$ or $c,c' \in L_{r,g}$.
    These are \emph{indifferences}, since $c$ and $c'$ are roughly indistinguishable w.r.t.\  $g$. 
\end{itemize}
Notice that these ordinal differences may be sensitive to the choice of anchor and candidates, hence one might ignore the distinction between weak and strong differences. However, since we compare our results to those of \citet{sigir21:filipandkristian}, who use this distinction, we do the same in this section.

The resulting \emph{SoftAttributes} data set consists of approximately 250K pairwise comparisons of movies over these 60 soft attributes and has a number of appealing properties. Of note is the fact that, since  each rater is asked to evaluate the same $60$ attributes, it does not suffer from tag popularity bias, in contrast to MovieLens-20M. However, subjectivity may be present in some attributes as raters may disagree on their usage.

Of the $60$ attributes used in the \emph{SoftAttributes} data set, only $36$ are found (frequently) in the MovieLens-20M data set.
When evaluating CAVs trained using MovieLens tags, we do so only on these 36 in-common attributes.
\footnote{
The in-common attributes are: 
animated,
artsy,
believable,
big budget,
bizarre,
boring,
cheesy,
complicated,
confusing,
dramatic,
entertaining,
factual,
funny,
gory,
harsh,
incomprehensible,
intense,
interesting,
long,
original,
over the top,
overrated,
pointless,
predictable,
realistic,
romantic,
scary,
unrealistic,
violent,
cartoonish,
exaggerated,
light-hearted,
mainstream,
mindless,
terrifying, and
sappy.
}

\subsection{Evaluation of CAVs using MovieLens Data}
\label{sec:ratereval_MovieLens}

We begin by evaluating how well CAVs, trained using the MovieLens-20M data set, can predict the assessments of \emph{SoftAttributes} raters. These CAVs are trained exactly as in the preceding sections, using collaborative filtering on ratings to generate user and item embeddings, and then using tag data to infer CAVs with our negative-label imputation scheme. We use the same procedures as above, namely, two-tower collaborative filtering and various methods for CAV training (logistic regression, RankNet and LambdaRank) using both straightforward objective methods and EM-based methods for sense subjectivity. The only difference with our earlier models is that our user/item embedding dimension is $d = 128$ in the collaborative filtering model. We use the larger embedding space to provide somewhat better recommendation performance and more nuance in the constructed item embeddings.

We evaluate the accuracy of our CAVs on each of the 36 in-common soft attributes/tags on the movies in the \emph{SoftAttributes} data set. Apart from evaluating CAVs using more precise human-rater data, this also provides an informal assessment of the degree to which CAVs trained on tag data generated by one user population under specific conditions generalize to attribute usage by a different population under different conditions.

Given some attribute $g$ with its CAV $\phi_g$, and an arbitrary pair of movies $(i,j)$ with item embeddings $\phi_I(i)$ and $\phi_I(j)$, respectively, the sign of
$\cos(\phi_I(i), \phi_g) - \cos(\phi_I(j), \phi_g)$ determines the CAV assessment of the relative degree of attribute $g$ for this pair, with a positive sign indicating $i \succ^{\phi_g}_g j$ and a negative sign indicating $j \succ^{\phi_g}_g i$.\footnote{We find that the use of cosine similarity gives a slightly better results than the use of dot products due the inherent normalization it provides. But results using dot products are qualitatively similar.}
Following \cite{sigir21:filipandkristian}, we compare these predictions against the rater assessments in \emph{SoftAttributes} using the \emph{extended Goodman-Kruskal gamma rank correlation coefficient}~\cite{goodman_kruskal_rank} $G'$. Specifically, for any attribute $g$, let a \emph{weak (resp., strong) disagreement} with the CAV $\phi_g$ be any $r,i,j$ triple where  $i \succ^{\phi_g}_g j$ but $j \succ^w_{r,g} i$ (resp., $j \succ^s_{r,g} i$). Weak and strong agreements are defined analogously. 
Let $N_s$ be the the number of weak agreements of $\phi_g$ with the data (w.r.t.\ $g$, we suppress $g$ in the notation), 
$N_{ss}$ the number of strong agreements,
$N_d$ the number of weak disagreements, and
$N_{dd}$ the number of strong disagreements. The extended Goodman-Kruskal gamma rank correlation coefficient (for a given attribute $g$) is defined as:
\begin{equation*}
G'_g = \frac{N_s - N_d + 2(N_{ss} - N_{dd})}{N_s + N_d + 2(N_{ss} + N_{dd})}\;.
\end{equation*}
Note that $G'_g$ is always between $-1$ (perfect anti-correlation) and $+1$ (perfect correlation). The ``double'' weight of strong agreements and disagreements is a simple extension of the Goodman-Kruskal metric adopted by \citet{sigir21:filipandkristian}. We use $G'$ to denote aggregate rank correlation across \emph{all attributes} where the agreement and disagreement counts are aggregates (effectively, a weighted average of $G'_g$ over all $g$). 

When building models that incorporate sense subjectivity, we apply our EM procedures as above, computing distinct CAVs for each uncovered sense of a specific attribute $g$. We limit the number of senses for any attribute $g$ to 10, but use a simple model selection criterion to select the precise number of senses. Specifically, we increase the number of senses from one (no subjectivity) by adding one additional sense at a time: if the improvement in average CAV quality when increasing from $k$ to $k+1$ senses falls below a fixed threshold, we terminate with $k$ senses, but stop at a maximum of $k=10$ senses if this threshold is not met at any stage. Since the users who apply tags in the MovieLens data differ from the raters evaluating attributes in the \emph{SoftAttributes} data set, we assign each \emph{SoftAttributes} rater to the attribute sense that best explains their ranking data for that attribute.

For reference, we also report the results of two methods proposed by \citet{sigir21:filipandkristian} for determining the semantics of soft attributes, weakly-supervised weighted dimensions+term based-item-centric (WWD+TB-IC) and
weakly-supervised weighted dimensions+term based-item-centric (WWD+TB-RC). While these methods are trained using different data sets than our use of MovieLens tags---they aggregate Amazon movie reviews joined with MovieLens~\cite{docent} to obtain attribute labels for movies---they are similar in these sense that they also use binary attribute labels rather than rater comparison data (as we discuss in the next section). Both methods fit a logistic regression model using item embeddings of dimension $d=25$. We refer to the original paper for further details.

\begin{table}[t]
    \centering
\begin{tabular}{|c||c|}
\hline
CAV Training Method & Gamma Rank Correlation \\ \hline\hline
WWD+TB-IC & 0.194~\cite{sigir21:filipandkristian} \\ \hline     
WWD+TB-RC & 0.200~\cite{sigir21:filipandkristian} \\ \hline
Logistic Regression, Lin.\ Emb. & 0.226 \\ \hline
EM Logistic Regression, Lin.\ Emb. & 0.336 \\ \hline
RankNet, Lin.\ Emb. & 0.199 \\ \hline
EM RankNet, Lin.\ Emb. & 0.388 \\ \hline
LambdaRank, Lin.\ Emb. & 0.191 \\ \hline     
EM LambdaRank, Lin.\ Emb. & 0.379 \\ \hline     
Logistic Regression, NL\ Emb. & 0.238 \\ \hline
EM Logistic Regression, NL\ Emb. & 0.329 \\ \hline
RankNet, NL\ Emb. & 0.216 \\ \hline
EM RankNet, NL\ Emb. & 0.339 \\ \hline
LambdaRank, NL\ Emb. & 0.203 \\ \hline     
EM LambdaRank, NL\ Emb. & 0.371 \\ \hline     
\end{tabular}
\vspace*{2mm}
   \caption{CAV Accuracy Evaluation, Training with MovieLens}
   \label{tab:training_with_movielens_tags}
\end{table}

Table~\ref{tab:training_with_movielens_tags} shows the performance of CAVs trained with MovieLens tag data using various techniques, specifically, the Goodman-Kruskal rank correlation with \emph{SoftAttributes} rater data. CAVs generated by all 12 methods examined exhibit a positive correlation with raters' attribute assessments of the 36 attributes, showing reasonable generalization of the semantics across the two different groups (and types) of users/raters. Our logistic regression methods (with both linear and nonlinear attribute embeddings) attain somewhat higher accuracy than both WWD+TB-IC and WWD+TB-RC. Apart from better performance by CAVs, this also suggests that the use of the tag data and our technique for negative data imputation has value for learning the semantics of soft attributes. Setting aside the EM methods designed for sense subjectivity for the moment, we see that each method trained using nonlinear embeddings achieves a slightly better result than its linear counterpart, a result consistent with those in Table~\ref{tab:non_subjective_movielens} above, suggesting that the preferences of raters for at least some soft attributes are nonlinear in their embedding-space representations.  

When assessing the role of subjectivity, we first note that (non-EM) logistic regression performs as well as our (non-EM)
ranking methods. This suggests that degree subjectivity may play a rather small role in rater assessments in the \emph{SoftAttributes} data set.\footnote{Note that the results for ranking methods vs.\ logistic regression in Table~\ref{tab:non_subjective_movielens} are based on different training/test sets.}
By contrast, the EM methods used to identify sense subjectivity do a significantly better job of predicting the attribute assessments of raters---overall, they seem to disentangle the distinct senses used by different raters, leading to much better rank correlation than objective (single-sense) methods. With EM, nonlinear embeddings and linear embeddings give similar results, suggesting some nonlinearity in attribute semantics can be captured by sense subjectivity.
\begin{table}[t]
\begin{tabular}{|c||c|c|c|c|c|c|} \hline
\multirow{2}{*}{CAV Training Method}& \multicolumn{6}{c|}{Number of Senses (Count, Average)}\\ \cline{2-7}
& 1 & 2 & 3 & 4 & 5 & Average \\ \hline\hline
EM Logistic Regression, Lin.\ Embs. & 21 & 7 & 5 & 0 & 3 & 1.806 \\ \hline
EM RankNet, Lin.\ Emb. & 4 & 18 & 8 & 6 & 0 & 2.444 \\ \hline
EM LambdaRank, Lin.\ Emb. & 5 & 14 & 12 & 3 & 2 & 2.528 \\ \hline
EM Logistic Regression, NL\ Embs. & 20 & 6 & 4 & 4 & 2 & 1.944 \\ \hline
EM RankNet, NL\ Emb. & 10 & 18 & 6 & 2 & 0 & 2.0 \\ \hline
EM LambdaRank, NL\ Emb. & 5 & 19 & 5 & 5 & 2 & 2.444 \\ \hline
\end{tabular}
\vspace*{2mm}
\caption{Number of CAV Senses Discovered by Various EM Methods (MovieLens 20M). Displayed for each method: (a) how many of the 36 attributes have been assigned the specified number of senses (1--5), and (b) the average number of senses over all 36 attributes.}
\label{tab:sense_stats_movielens}
\end{table}
Table~\ref{tab:sense_stats_movielens} shows the number of senses discovered by the EM methods across all 36 attributes.  We see that logistic regression with EM rarely uncovers more than one sense for a majority of attributes---21 (resp., 20) of the 36 attributes have a single sense in the case of linear (resp., nonlinear) CAVs. and 20 with nonlinear embedding). On average, the EM ranking methods do discover a greater number of senses than logistic regression models.

In Table \ref{tab:negative_sampling_comparison} we show the results of an ablation study to test the effect of negative sampling on various CAV training methods. When training the CAV for a specific attribute, we select from one to five negative exemplars randomly for each positive example (where negatives are defined as in Section~\ref{sec:linear}).  We see that results are not especially sensitive to ratio of negative to positive examples. With logistic regression, 2--4 negatives works best providing an improvement of about 0.03 (resp., 0.05) in gamma rank correlation for the non-subjective (resp., subjective) case. Our ranking methods show negligible improvement when adding more than one negative sample. We suspect this is so because logistic regression is trying to separate negative and positive examples, thus additional negative samples assists the model in finding a better classification boundary, though too many puts too much weight on the noisy negative samples relative to ground-truth positive examples. On the other hand, in ranking only the relative ordering matters, so additional negative samples (which are inherently more noisy since they are imputed) adds little value and may even be detrimental. Given these results, we use a negative sampling rate of 4:1 for logistic regression and 1:1 for ranking in all following experiments.

\begin{table}
\begin{tabular}{|c||c|c|c|c|c|c|}
\hline
No. Negative Samples & LogRegr & EM LogRegr & RankNet & EM RankNet & LambdaRank & EM LambdaRank \\ \hline
1 & 0.205 & 0.288 & 0.199 & 0.388 & 0.191 & 0.379 \\ 
2 & 0.234 & 0.319 & 0.223 & 0.390 & 0.195 & 0.370 \\
3 & 0.213 & 0.331 & 0.207 & 0.359 & 0.211 & 0.346 \\
4 & 0.226 & 0.336 & 0.197 & 0.382 & 0.215 & 0.342 \\
5 & 0.218 & 0.307 & 0.213 & 0.338 & 0.194 & 0.340 \\ \hline 
\end{tabular}
\vspace*{2mm}
    \caption{Comparison of negative sampling with different numbers of samples for the linear embedding case.}
    \label{tab:negative_sampling_comparison}
\end{table}

\subsection{Evaluation of CAVs Trained using \emph{SoftAttributes} Rater Data}
\label{sec:ratereval_SA}

In contrast to the evaluation above---where CAVs were trained using MovieLens tags and used to predict the attribute assessments of \emph{SoftAttributes} raters---we now train CAVs directly on rater assessments of soft attributes and evaluate their predictive quality. Since rater assessments are ordinal rather than binary, we train CAVs using RankNet and LambdaRank, but not logistic regression.

We train and test the CAV $\phi_g$ for attribute $g$ using the pairwise comparisons induced by each rater's assessment of $g$, specifically, applying RankNet and LambdaRank to the data set consisting of per-rater pairwise comparisons. In training, we set the instance weight to two for comparisons with strong differences and one for weak differences.
When considering sense subjectivity, we use EM variants of our two approaches to ``cluster raters'' based on their usage as we do above when training with MovieLens data. The number of senses for any specific $g$ is selected in the same fashion as above, using a specific threshold on performance to stop increasing the number of different senses.
We do not use nonlinear embeddings in this assessment, since results above (see Table~\ref{tab:training_with_movielens_tags}) suggests that nonlinear embeddings provide only marginal improvements over linear embeddings when coupled with sense subjective modeling (or EM). We also report the results of the Supervised Weighted Dimensions (SWD) method used by \citet{sigir21:filipandkristian} for reference. SWD uses a linear ranking support vector machine~\cite{joachims02} induced using the same raters' pairwise comparisons as those used to train CAVs (we refer to the original paper for additional details).

We report CAV results using a 5-fold testing procedure. Specifically, to evaluate the performance of each CAV method on attribute $g$, we split raters into five groups of 20 at random, train the CAVs on each subset of four groups (80\% of the data) and evaluate the CAV on the fifth (20\% of the data), and average results over the five train/test splits---we do this to account for the potential noise in using only 20 raters for testing. For evaluation, we use extended Goodman-Kruskal gamma rank correlation to assess how well CAV predictions conform to the item rankings of test raters. As above we report the aggregate  coefficient $G'$ (i.e., weighted average results over all 60 attributes).

\begin{table}[t]
    \centering
\begin{tabular}{|c||c|}
\hline
CAV Training Method & Gamma Rank Correlation \\ \hline\hline
SWD & 0.485~\cite{sigir21:filipandkristian} \\ \hline     
RankNet, Lin.\ Emb. & 0.523 \\ \hline
EM RankNet, Lin.\ Emb. & 0.667 \\ \hline
LambdaRank, Lin.\ Emb. & 0.522 \\ \hline     
EM LambdaRank, Lin.\ Emb. & 0.672 \\ \hline     
\end{tabular}
\vspace*{2mm}
   \caption{Evaluation of Accuracy of CAVs Trained using \emph{SoftAttributes} data set. Average Aggregate $G'$ (extended Goodman-Kruskal gamma rank correlation) with test set (5-fold).}
   \label{tab:training_with_softattributes}
\end{table}

\begin{table}[t]
\begin{tabular}{|c||c|c|c|c|c|} \hline
CAV Training Method & [2.5, 3.5) & [3.5, 4.5) & [4.5, 5.5) & [5.5, 6.5) & Average Number of Senses \\ \hline\hline
EM RankNet, Lin.\ Emb. & 11 & 31 & 18 & 0 & 4.153 \\ \hline
EM LambdaRank, Lin.\ Emb. & 10 & 26 & 21 & 3 & 4.223 \\ \hline
\end{tabular}
\vspace*{2mm}
   \caption{Number of CAV Senses Discovered by Various EM Methods (\emph{SoftAttributes} data set). Displayed for each method: (a) how many of the 60 attributes have been assigned the specified number of senses---since we use 5-fold testing, we use the \emph{average} number of senses across the five test sets for each attribute (bucketed around integer values 3 through 6); and (b) the average number of senses over all 60 attributes.}
   \label{tab:sense_stats_softattributes}
\end{table}

Table~\ref{tab:training_with_softattributes} shows the performance of CAVs trained on the \emph{SoftAttributes} data set.
The CAV methods show highly positive correlation between the learned CAVs and rater assessment of attibute values. This suggests that the ability to learn the semantics of soft attributes directly from ground-truth user comparison of a small set of items of w.r.t.\ the attribute in question can dramatically improve results relative to CAVs trained indirectly using tags (see Table~\ref{tab:training_with_movielens_tags}). Both RankNet and LambdaRank outperform SWD, while the difference between RankNet and LambdaRank is minimal. 
The EM ranking methods outperform non-EM techniques, suggesting that sense subjectivity is a major factor in rater assessments in the \emph{SoftAttributes} data, though the difference between EM RankNet and EM LambdaRank is negigible.
Table~\ref{tab:sense_stats_softattributes} shows the number of distinct senses (averaged over 5-fold train/test) uncovered by the EM methods. The two methods differ very little, EM LambdaRank discovering only slightly more senses than EM RankNet.

\section{Using CAVs for Example Critiquing}
\label{sec:critiquing}

While preference elicitation in recommender systems often uses attributes \cite{pu:AIM2008,viappiani:nips2010,bonilla_sanner:nips10}, an important question is the extent to which such methods can 
be adapted to handle soft attributes (see, e.g., \cite{radlinski:sdd2019}).
The sheer variety of approaches to preference elicitation means we cannot consider this question in its full depth in this work. Instead, we focus on one particular methods by which users can express their preferences explicitly in recommenders, namely \emph{example critiquing} \cite{chen_critiquing_survey:umuai2012},
and examine how the CAV semantics for tags can be used for this purpose.
In lieu of live experiments, we adopt a stylized but plausible \emph{user response model} in which a user's critiques are driven by her underlying utility function and her personal tag/attribute semantics.\footnote{Synthetic user models are commonly used to evaluate recommender systems \cite{zou:neurips20,aliexpress:tkde21}.}
While other response models are possible (e.g., models that are fit to real user interaction data), this model suffices to demonstrate the value of CAVs for critiquing. We run synthetic experiments in which we have access to the ground truth utility and semantics for each user, and a MovieLens experiment, for which we propose a novel method for generating utilities and responses.

\vskip 2mm
\noindent
\textbf{Recommender System Interaction Model.} \hspace*{2mm}
We assume a predefined list of critiquable tags and adopt a simple interactive recommender system that supports user critiques. We provide a brief description of our set up here but refer to Appendix~\ref{app:critiquing} for further details, parameter values used, etc. In the interactive system, each interaction with user $u$ has the recommender present a slate $S$ of $k$ items to $u$. User $u$ can \emph{accept} one of the recommended items, at which point the session terminates. Otherwise, $u$ can \emph{critique} $S$ using a tag $g$ and a desired direction (`more,' `less'). For instance, given a slate of $k$ movies from which to choose, $u$ might offer a critique like ``more funny'' or ``less violent,'' intending that the recommender system makes an improved slate of next recommendations at the next iteration that reflects this preference. The recommender system then updates its \emph{user representation} given this response and generates the next recommended slate. The process repeats until the user terminates by accepting a recommendation or the recommender system reaches the maximum critiquing steps $T$. We describe below the processes by which the recommender selects slates and updates its user representation, and by which the user responds to recommended slates.

\vskip 2mm
\noindent
\textbf{User Response Model.} \hspace*{2mm}
Each behavior is driven by an underlying user model which is identical in form to the models used in our synthetic generative model above. Specifically, a user $u$ has (i) a (personal) ground truth utility function over items (i.e., $u$ can assess the utility of any item $i$ that is recommended) and (ii) a (personal) ground-truth semantics for attributes (i.e., $u$ can assess, for any recommender item $i$, $i$'s value of the soft attribute corresponding to any of the critiquable tags).
User interactions assume a \emph{user response model}.
User $u$ also has a rough estimate of the maximum and minimum levels any tag/attribute can attain in the item corpus---this information is used to guide her critiquing behavior without assuming she has unrealistic knowledge of the item corpus (see Appendix~\ref{app:critiquing}). When presented with slate $S$, $u$ accepts an item $i$ if its utility is sufficiently large, i.e., exceeds some threshold.
Otherwise, $u$ critiques slate $S$ using the \emph{most salient tag} $g$ with respect to utility improvement of $S$; in particular, $u$ critiques using the tag $g = \argmax_g \delta_u^T w_g $, where $$\delta_u=(\phi_I(i^*_u)-\frac{1}{|S|}\sum_{i\in S}\phi_I(i))\odot \phi_U(u)$$
is the utility difference vector between user $u$'s estimated ideal item $i^*_u$ and her average utility vector over the $k$ items in $S$, and $w_g$ is $u$'s interpretation of $g$. $\odot$ refers to the Hadamard product.

\vskip 2mm
\noindent
\textbf{Recommendation Strategy.} \hspace*{2mm}
We assume item embeddings $\phi_I(i)$ are fixed and use a \emph{simple heuristic recommendation strategy} for incorporating critiques. We emphasize that the recommendation strategy used and method for incorporating critiques are fairly generic and are not intended to reflect the state-of-the-art, since our goal is to measure the ability to exploit learned CAVs. More elaborate strategies for updating user embedding are possible, including the use of Bayesian updates relative to a prior over the user embedding \cite{vendrov2020gradient}. Here instead we adopt a simple heuristic, based on \cite{luo2020latent}, to focus attention on the CAV semantics itself.

Given a user embedding $\phi_U(u)$, the recommender system scores all items $i$ in the corpus with respect to utility $r_{i,u} = \phi_I(i)^T\phi_U(u)$, and presents the slate $S$ of the top $k$ scoring items. 
If the user critiques $S$ with a tag $g$ (and a specific direction), the system updates the user embedding as follows:
$$\phi_U(u) \leftarrow \phi_U(u) + \alpha_t(g) \cdot \phi_g$$, where $\phi_g$ is $g$'s CAV, $\alpha_t(g)$ is a tag-specific step size at iteration $t$ (the $t$th interaction with $u$), and the sign of $\alpha_t(g)$ reflects the direction (`more,' `less') of the critique. We treat the magnitude of $\alpha_t(g)$ as a hyper-parameter selected to optimize the utilities of the resulting recommendations.\footnote{Ultimately, this heuristic adjustment should be tuned to the specifics of a user response model generated from live experiments.} The system then recommends the top $k$ items given the updated user embedding, and the process repeats. If the tag $g$ is sense-subjective, the critique is interpreted relative to the systems's estimate of $u$'s sense (or cluster) based on past usage.

\vskip 2mm
\noindent
\textbf{Synthetic Data set Results.} \hspace*{2mm}
We first analyze critiquing with CAVs using the same synthetic models described above, where a user's critiques are generated with the user's ground truth utility and tag semantics. By contrast, the recommender system uses its \emph{estimated user embedding} and the \emph{tag's CAV} to interpret a user's critique (and not the ground truth). In our experiment, the recommended slate size is $k=10$, and the maximum number of critiquing steps is $T=25$. We evaluate how recommendation quality improves as the number of critiques increases by measuring the user's utility for the sequence of recommended (top-$k$) slates. We measure this ``slate utility'' using two different measures:
\begin{itemize}
\item \emph{User max utility} of the top-$k$ slate: $\UMU(S) = \max_{i\in S} U(i)$, where $U(i)$ is $u$'s \emph{true} utility for $i$. This reflects the utility of a user who is able to select their most preferred item from $S$.
\item \emph{User average utility} of the top-$k$ slate: $\UAU(S) = \mathrm{avg}_{i\in S} U(i)$. This captures the utility of a user who randomly selects an item from the recommended slate.
\end{itemize}

\begin{figure}
\centering
\captionsetup{width=.45\linewidth}
\begin{minipage}{.5\textwidth}
  \centering
\includegraphics[width=.95\linewidth]{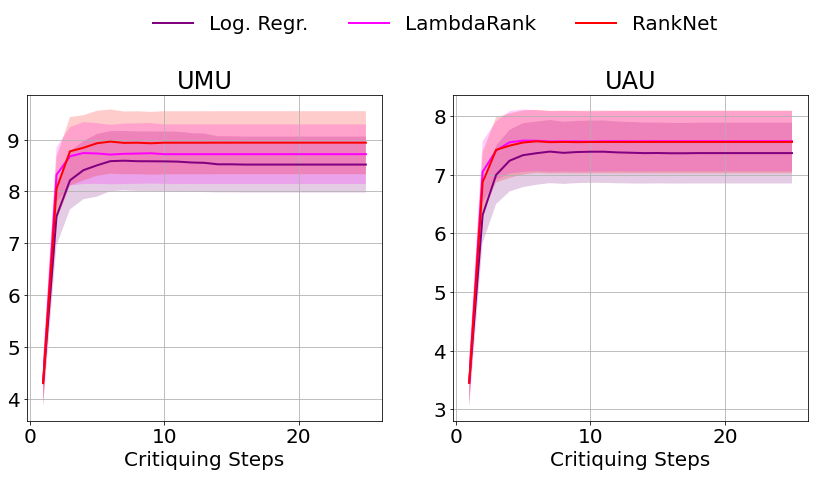} \\
\includegraphics[width=.95\linewidth]{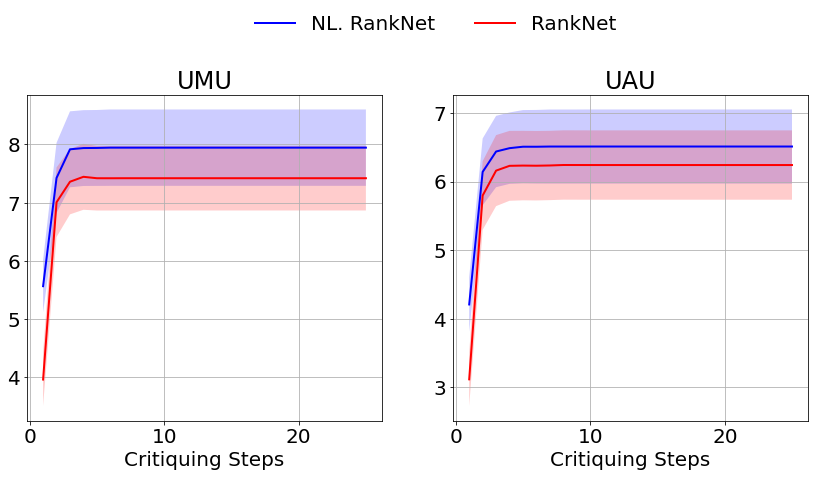} \\
\includegraphics[width=.95\linewidth]{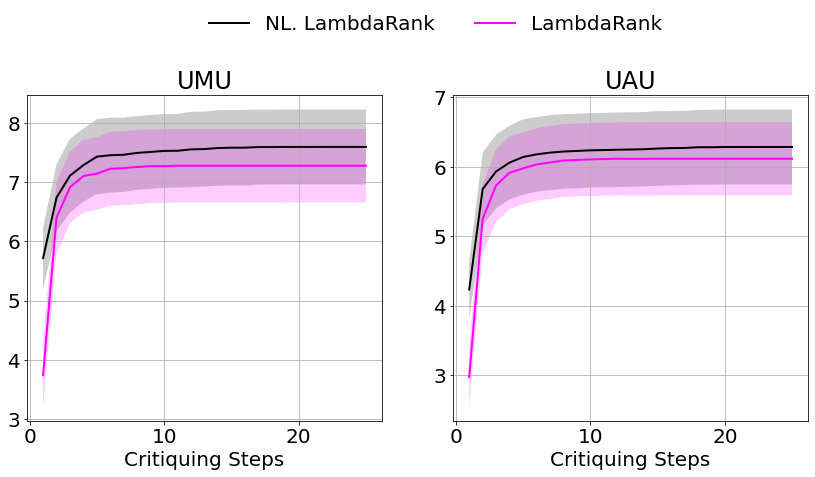} 
  \vspace{-0.1in}
\caption{Results of Interactive Critique with Synthetic Data. Top two: No Subjectivity Linear Utility with All Methods; Middle two: No Subjectivity Nonlinear Utility with RankNet; Bottom two: Degree Subjectivity Nonlinear Utility with LambdaRank}
  \vspace{-0.1in}
\label{fig:critique1}
\end{minipage}%
\begin{minipage}{.5\textwidth}
  \centering
\includegraphics[width=.95\linewidth]{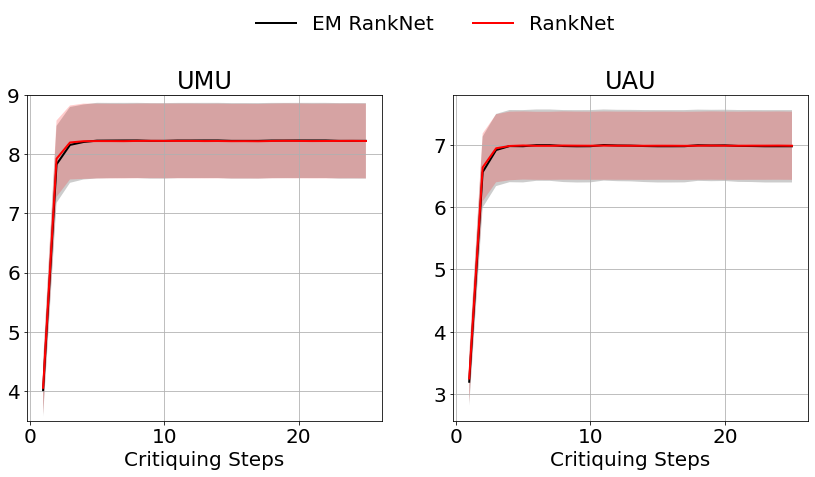} \\
\includegraphics[width=.95\linewidth]{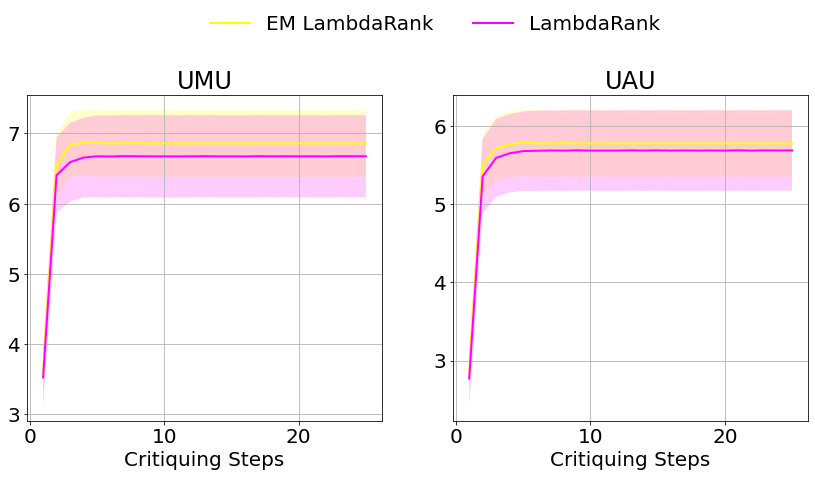} \\
\includegraphics[width=.95\linewidth]{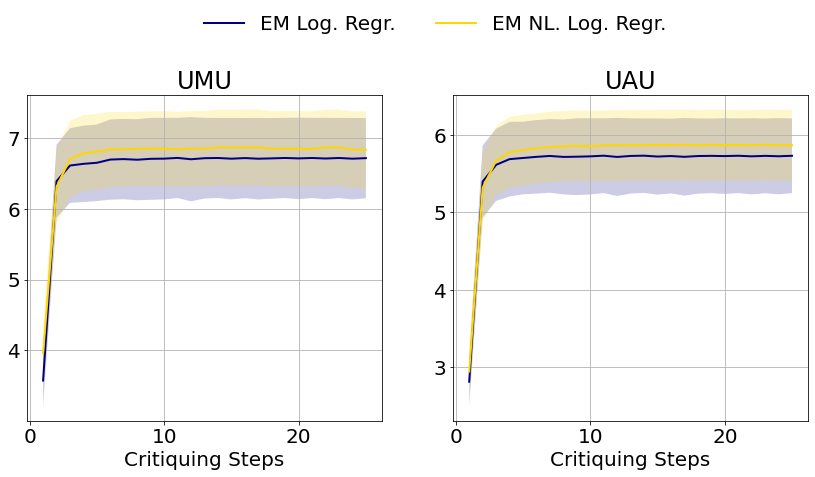}
  \vspace{-0.1in}
\caption{Results of Interactive Critique with Synthetic Data. Top two: Sense Subjectivity Linear Utility with RankNet; Middle two: Sense Subjectivity Nonlinear Utility with LambdaRank, Lin-Emb; Bottom two: Sense Subjectivity Nonlinear Utility with Log. Regr., NL-Emb.}
  \vspace{-0.1in}
\label{fig:critique_subjective_sense}
\end{minipage}
\end{figure}

Fig.~\ref{fig:critique1} presents the interactive critiquing results from three experiments using different synthetic data sets. The first has no subjectivity in user tag usage and assumes linear utility. The second also uses no subjectivity, but user item utility is nonlinear in the soft attributes underlying the tags. Finally, the third allows both degree subjectivity and nonlinear utility. We see from Fig.~\ref{fig:critique1} that in all experiments, user utility, both $\UMU$ and $\UAU$, improves with additional critiquing steps, eventually converging to a steady-state value. These results corroborate our hypothesis that, since CAVs quite accurately represent soft attributes in embedding space, they can be used to effectively update the recommender system's beliefs about user preferences as the user critiques recommended items, which in turn improves recommendation quality.

Furthermore, we recall from
Sections~\ref{sec:objectiveCAVs} and~\ref{sec:subjectiveCAVs}
that CAVs trained with logistic regression generally have lower accuracy than those trained with RankNet or LambdaRank. 
While our CAV algorithms are not optimized to support critiquing, more accurate CAVs learned using ranking methods give rise to better interpretations of user critiques by the recommender---this observation is reflected by both the faster improvement and the greater steady-state values for both $\UAU$ and $\UMU$ in our experiments. 
This is most likely due to the fact that more accurate ranking-based CAVs better capture a user's intended semantics during critiquing. Similarly, the improved accuracy of nonlinear CAVs when utility is nonlinear manifests in the improved critiquing performance (in both the objective and degree-subjectivity tests, and for both $\UAU$ and $\UMU$).
Again, this performance improvement is likely due to the better CAV representation uncovered from the intermediate layers of the DNN.

Figure~\ref{fig:critique_subjective_sense} shows interactive critiquing results using synthetic data in which the critiquable tags exhibit \emph{sense subjectivity} under three different experimental settings: (i) user utility is linear w.r.t.\ soft attributes; (ii) user utility is nonlinear but the trained CAVS are linear CAVs; and (iii) both user utility and the trained CAVs are nonlinear. As above, both $\UMU$ and $\UAU$ improve with a increasing number of critiquing steps, eventually converging to a steady-state value. Recall that only three of the 25 utility-relevant dimensions correspond to the ``conflated'' senses of a single tag. Still, in the case of nonlinear utility, EM-LambdaRank, even with linear CAVs, outperforms LambdaRank without EM. This suggests that disentangling sense subjectivity is more critical than having an ``accurate'' but incorrectly-assumed objective single CAV.
These results also show that EM Logistic Regression can be improved with the use of nonlinear CAVs as suggested in Table~\ref{tab:subjectivity_sense}.

\vskip 2mm
\noindent
\textbf{MovieLens20M Results.} \hspace*{2mm}
To evaluate critiquing with soft attributes using the MovieLens20M data set, we propose a novel method for hypothesizing ``ground-truth'' user utility. We first train a collaborative filtering model with all users and items, then train (non-subjective) CAVs for $164$ tags (as described in Section~\ref{sec:empiricalObjective}). We then construct a small set of \emph{test users}, each of whom has rated at least $50$ movies. We use the learned embedding $\phi_U(u)$ for each test user $u$ as if it were their \emph{ground truth utility}. Since $u$ has rated a large number of movies, we expect this ground truth utility to be reasonably stable and accurate. We then run the interactive critiquing recommender system by \emph{forgetting} each test user---i.e., forgetting both their ratings and tags---and treating them as a ``cold start'' user, who is given a generic prior embedding.\footnote{We use the average of all learned user embeddings as a prior.} This user $u$ then generates critiques of the recommended slates using $\phi_U(u)$ as their true utility. Since we have no ground truth tag semantics, each $u$ treats the recommender system's \emph{learned CAV} as her semantics (admittedly giving the recommender system some advantage when interpreting critiques). Otherwise, the user response model is exactly as in the synthetic case. We evaluate as in the synthetic case, but user utility improvements are \emph{estimates} using the learned embedding $\phi_U(u)$. Because of this we also assess some additional metrics (see below).

Fig.~\ref{fig:critique_movielens} shows critiquing results with MovieLens data. In addition to $\UAU(S)$, we also report normalized discounted cumulative gain (NDCG)  \cite{valizadegan2009learning}, mean reciprocal rank (MRR) \cite{mcfee2010metric}, and average binarized rating\footnote{We set the binarized rating to $1$ if the numeric rating exceeds $3$, and $0$ otherwise. This is a binary precision measure with respect to highly-rated items.} \cite{rashid2002getting} of slates $S$ generated during critiquing. We compare critiquing results generated by four sets of CAVs, trained with the following methods: RankNet, nonlinear RankNet, nonlinear LambdaRank, and nonlinear logistic regression (as these methods usually generate more accurate CAVs as discussed above).
While all methods perform similarly with respect to $\UAU$, nonlinear LambdaRank outperforms the others with respect to the other three metrics. This again provides evidence that: (i) incorporating CAVs in recommender systems to capture user-critiquing behavior can improve recommendation quality (both utilities and ratings); and (ii) the performance of critiquing-based recommender systems tends to improve with the quality of the learned CAVs.

\begin{figure}
    \centering
\includegraphics[width=1.\textwidth]{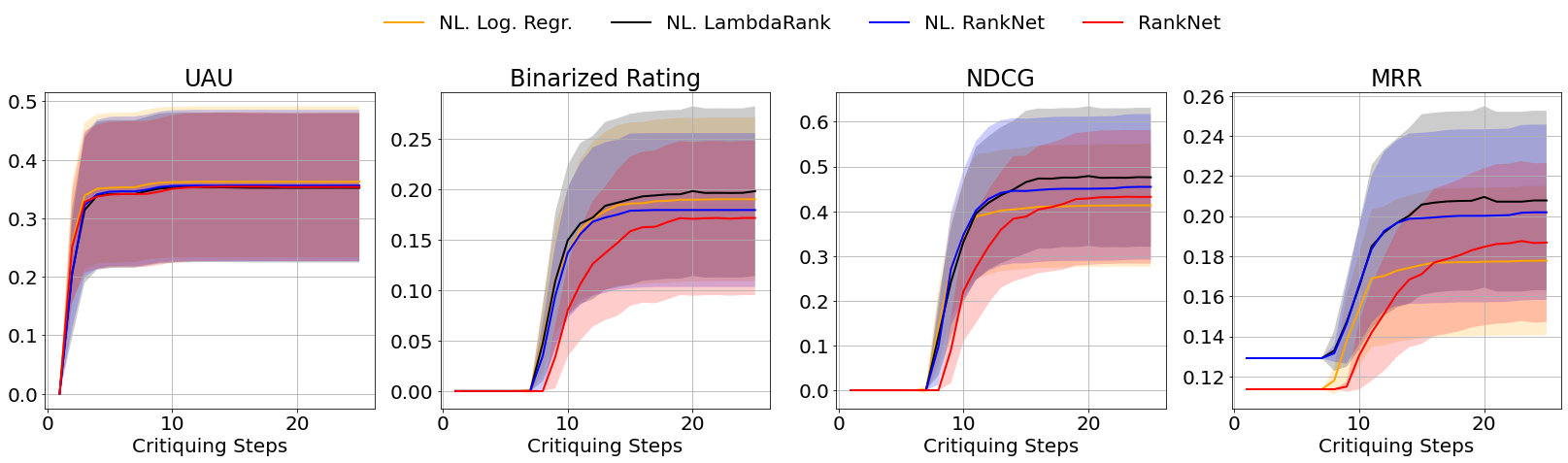}
  \vspace{-0.1in}
\caption{Results of Interactive Critique with MovieLens Data}
  \vspace{-0.1in}
\label{fig:critique_movielens}
\end{figure}

\section{Conclusions}
\label{sec:conclude}

We have presented a novel methodology for discovering the semantics of soft attribute/tag usage in recommender systems using concept activation vectors. Its benefits include: (i) using a collaborative filtering representation to identify attributes of greatest relevance to the recommendation task; (ii) distinguishing objective and subjective tag usage; (iii) identifying personalized, user-specific semantics for subjective attributes; and (iv) relating attribute semantics to preference, thereby allowing interactions using soft attributes/tags in example critiquing and other forms of preference elicitation.

A number of future directions can provide additional value to the use of CAVs for elicitation and critiquing in recommender systems. While our empirical results suggest that CAVs are useful for subjective attributes and critiquing, additional study with real user critiques is critical, as is developing real-world data sets with ground truth utility and personal semantics (e.g., via survey instruments or controlled experiments).

In our current formulation, CAVs are learned offline under the assumptions that the data set contains either sufficient tagging data reflecting the personal preferences and semantics of diverse users 
or results of \emph{pairwise tag-comparison queries} from sufficiently diverse users.
When this is not that case, a useful extension is to consider an interactive methods in which the system can actively elicit a user's personal semantics, for example,
using \emph{pairwise tag-comparison queries} to identify a user's tag interpretation. For example, a few well-chosen queries based on offline data like
``Which of these two books is more thought-provoking?'' or ``Do you consider this song to be upbeat?'' could help identify a user's personal semantics (both degree thresholds and tag senses).

Our focus in this work has been on soft attribute usage in an item-space that is characterized entirely by latent attributes (e.g., as is common in content recommendation). Other domains, such as product recommendation, are traditionally characterized by hard attributes (e.g., price, color, weight, efficiency, capacity, and other product-specific features). However, the use of CAVs to allow the more flexible use of soft and subjective attributes (e.g., cozy, comfortable, vibrant, stylish, compact, inexpensive, etc.) to navigate such domains is of great interest. This raises interesting questions of how to integrate soft and hard attributes, including the possibility of exploiting the ground-truth semantics of hard attributes to facilitate the discovery of soft-attribute semantics (e.g., relating (possibly subjective) terms like ``compact'' to product dimensions or weight, or ``inexpensive'' to price).

Finally, it would be of great value for critiquing-based recommender systems to have
CAV-learning algorithms that not only maximize concept accuracy, but whose representations are more directly tuned to support preference elicitation.
and to more directly compare CAVs to alternative ways of understanding soft and subjective attributes.

\section*{Acknowledgements}

Thanks to Krisztian Balog, Walid Krichene, Dima Kuzmin, Nicolas Mayoraz, Filip Radlinski, Steffen Rendle, Tania Bedrax-Weiss, and Li Zhang for helpful discussions and feedback, as well as the anonymous reviewers of an earlier version of the manuscript.

\bibliographystyle{ACM-Reference-Format}
\bibliography{long,standard,tcav}

\appendix

\section{Appendix}

We provide additional details on various aspects of this work.
We fully specify the synthetic data generation process used in App.~\ref{app:syntheticdata}, describe our precise use of the MovieLens data set in 
App.~\ref{app:movielens},  elaborate on our training methods (incl.\ model architectures, parameter values, etc.) in App.~\ref{app:training}, and provide addition detail on our example critiquing set up in App.~\ref{app:critiquing}.

\subsection{Synthetic Data Generation with RecSim}
\label{app:syntheticdata}

We use a stylized, but structurally realistic generative model to produce synthetic ratings and tag data for some of our experiments. Its purpose is twofold. First, it offers various parameters or ``knobs'' that can be used to generate data sets that can increase/decrease the level of difficulty faced by methods designed to extract the semantics of soft or subjective attributes. Second, synthetic data generation provides us with a ``ground truth'' against we can test (i) the quality of our learned CAV representations of soft attributes and (ii) the effectiveness of our elicitation methods at using soft, subjective attributes to improve recommendations.

The generative user-response model  is implemented using RecSim NG \cite{mladenov2020recsimng}, a platform for simulating user behavior when interacting with recommender systems that supports the authoring of structured, graphical and causal models of agent behavior or learning such behavior from data. (We use the former capability.)

We start by describing the process for generating ratings and tags for ``non-subjective'' tags, where users have linear utility for the corresponding soft attributes. We then describe mild modifications of this core generative model to allow for subjective tags (both degree and sense subjectivity) and nonlinear attribute utility.

\vskip 2mm
\noindent
\textbf{Non-subjective, linear utility model.} \hspace*{2mm}
The generative process proceeds in the following stages: we first generate items (with their latent and soft attribute values and other properties); then users (with their utility functions and other behavioral characteristics); then user-item ratings; and finally user-item tags. The model reflects realistic characteristics such as item and user ``clustering,'' popularity bias, not-missing-at-random ratings, the sparsity of ratings, the relative sparsity of tags compared to ratings, etc.

Specifically each item $i$ is characterized by an \emph{attribute vector} $\bfv(i)\in [0,1]^D$, where $D = L+S$: $L$ dimensions correspond to latent item features and $S$ to soft attributes. For a soft dimension $L < s \leq L+S$, $v^s(i)$ captures the degree to which $i$ exhibits soft attribute $s$. We sample $m$ items from a mixture of $K$ $D$-dimensional Gaussian distributions (truncated on $[0,1]^D$) $\calN(\mu_k, \sigma_k)$, $k\leq K$, with mean vector $\mu_k \in [0,1]^D$ and (diagonal) covariance $\sigma_k$. We set $D$ to 25 and $K$ to 100 in our experiments. For simplicity, we assume all $\sigma_k$ are identical to 0.5, and sample means uniformly. Mixture weights are sampled uniformly at random and normalized.
For each item $i$, we also randomly generate a popularity bias $b_i$ from $[0,1]$ (whose purpose is described below).

Each user $u$ has a \emph{utility vector} $\bfw(u)\in [0,1]^D$ reflecting its utility for items. We sample $n$ users from the above $K$-mixture-of-Gaussian distribution similar to that for items. Here the means and variances of these distributions are same as the ones used for item distributions,
but the mixture weights are fully resampled. This step ensures that the generated samples of users and items are distributed in different parts of the latent ``topic space''.

Next, we generate the user-item ratings with the following steps:
\begin{enumerate}[(i)]
\item For each $u$, we draw $\NumR_u$ samples from a Zipf (or zeta) distribution with a power parameter $a = 1.05$ to reflect the natural power law over the number of ratings provided by users. Parameter $a$ is chosen in a way that the average number of rated items by each user is approximately $100$.
We also set the maximum number of ratings by each user to $1,000$. 
\item To generate the candidate items to be rated by each user $u$, we
generate a set of $\Rated_u$ items, by sampling them without replacement from the overall set of items via a multinomial logit (or softmax) choice model, where the probability associated with each item $i$ is proportional to $e^{\tau \cdot (\bfw_u\bfv_i + b_i)}$. Here $\tau$ is a temperature parameter that controls the degree of randomness in choosing the item wrt greatest affinity, and we use $\tau = 1$ in our experiments.
\item For each user $u$ and item $i\in\Rated_u$, the rating $r_{ui}$ is generated as follows. We denote by $s(u,i):=\bfw_u\bfv_i + \veps$ be the score of item $i$, where $\veps$ is a small, zero-mean random noise. We then discretize all the scores provided by user $u$ into $5$ equally sized sub-intervals in $[\min_u, \max_u]$, where $\min_u = \min \{s(u,i): i \in\Rated_u\}$ and $\max_u = \max \{s(u,i): i \in\Rated_u\}$ and assign a $1$ to $5$ rating to each item  $i\in\Rated_u$ accordingly.
\end{enumerate}

For each soft attribute $s$ we assume there is a unique tag $g_s$ that users can apply when referring to that attribute. For each generic tag $g$, we denote by $s(g)$ the corresponding soft attribute (so $g = g_{s(g)}$). To complete the data-generation procedure we also generate user-item tags with the following steps. 
\begin{enumerate}[(i)]
\item For each user $u$, we generate $\PT_u$, the probability of tagging an item, from a mixture of two distributions: (a) a Dirac distribution at $0$ with weight $0<x<1$; and (b) a Uniform distribution
over $[p_-, p_+]$, where $0 < p_- \leq p_+$, with weight $1-x$. This reflects the fact that a large fraction of users never use tags, and among those who do, some users tag much more frequently than others. In our experiments, we set $x$ to 0.8, $p_-$ to 0.1, and $p_+$ to 0.5.

\item For each user $u$ with non-zero tag probability, i.e., $\PT_u > 0$, we first generate the set $\Tagged_u$, which represents the items that are tagged by $u$. Here $\Tagged_u$ is a subset of rated items $\Rated_u$ such that each rated item will be tagged with (independent) probability $\PT_u$. This reflects the fact that a user will not tag an unrated movie, but may leave some rated movies untagged.\footnote{One could also allow, if desired, the propensity to tag to vary with the tag $g$, and/or bias the application of tags to higher-rated items.} For any item $i\not\in\Tagged_u$, the corresponding indicator value $t_{u,i,g} = 0$ for every tag $g$, which means that no tag is applied by user $u$ on item $i$.

\item For every (non-subjective) tag $g$, we use a user-independent threshold $\tau_g = 0.5$ indicating the degree to which an item must possess attribute $s(g)$ to be tagged with tag $g$ by a user. 

\item For every item $i \in\Tagged_u$ and tag $g$, we set the indicator value $t_{u,i,g} = 1$ (i.e., user $u$ applied tag $g$ to item $i$) if $v^{s(g)}(i) \geq \tau_g + \veps$ (where $\veps$ is a small, zero-mean random noise drawn independently from $\calN(0, 0.01)$ for each $(u,i,g)$). Otherwise the indicator value $t_{u,i,g}$ remains at $0$. 

\end{enumerate}

\vskip 2mm
\noindent
\textbf{Subjective model.} \hspace*{2mm}
The generative model above is modified in a straightforward way to handle subjective attributes. For degree subjectivity, we generate user-dependent tag-application thresholds $\tau^u_g$ for each user-tag pair, rather than user-independent thresholds $\tau_g$. In general, to allow for some ``commonality'' across user sub-populations, we draw these thresholds from mixture distributions with a small number of components and small variance.  Threshold distributions that are widely dispersed can be viewed as ``fully'' subjective, while for objective attributes their thresholds (as above) are special cases for which each of their distributions follow a Dirac at $\tau_g$. In our experiments, for each degree-subjective tag the threshold is randomly chosen between $0.5$ and $0.7$.

For sense subjectivity, the model is as above with the following modification. We maintain $S_{\obj}$ soft attributes of the form above---which we now call \emph{objective}---each $s\in S_{\obj}$ corresponds to one item dimension and to a specific \emph{objective (in sense)} tag $g_s$. In addition, we have $S_{\subj}$ \emph{subjective} soft attributes, partitioned into \emph{tag groups}, $S^1, \ldots, S^J$ under the following condition: (a) $S^i \cap S^j =\emptyset$ for $i\neq j$; (b) $\cup_{j\leq J} S^j = S_\subj$; and (c) $|S^j| > 1$ for all $j\leq J$. Each tag group $S^j$ is associated with a single tag $g^j$, with each $s\in S^j$ reflecting a different \emph{sense} for $g^j$.

For each tag group $S^j$, each user $u$ is randomly assigned to exactly one such sense $s(u,j) \in S^j$. This has two implications. First, when user $u$ considers applying tag $g^j$ to an item, it is evaluated according to soft attribute $s(u,j)$. This means that $u$ uses that specific sense when applying that tag. Second, the utility vector $\bfw(u)$ of user $u$ is such that its
$s^{\textrm th}$ component is zero for each $s\in S^j$ except for $s(u,j)$. This implies that $u$ assesses her utility for an item using only her designated attribute (or sense) from each of the tag groups.

\vskip 2mm
\noindent
\textbf{Nonlinear utility model.} \hspace*{2mm}
There are many forms of nonlinearity. We propose one especially simple form that allows one to test whether our CAV method (or any other soft-attribute method) can identify such attributes. Specifically, we consider the simple single-peaked utility function, and for simplicity we only describe the  utility function for non-subjective attributes. Extending it to the subjective case is straightforward. 

Assume the domain of any arbitrary attribute $1 \leq a \leq D$ is $[0,1]$, and the user utility is \emph{additive-independent} across attributes, i.e., the utility for an item is the sum of ``local utilities'' for each attribute (the dot-product model satisfies this trivially). 
A user $u$’s utility function is \emph{single-peaked} with respect to $s$ if $u$ has an ideal point $p_{u,a} \in [0,1]$ such that $u$’s local utility for the attribute $l_{u,a}(x)$ is maximized at $p_{u,a}$ and decreases monotonically as $x$ moves away from $p_{u,a}$. The functional form of a single-peaked utility can be arbitrary with the simplest form being a piecewise linear function. Here we use $l_{u,a}(x) = p_{u,a} - |x - p_{u,a}|$, where $x = \bfw_{u, a}\bfv_{i, a}$. As both vectors $\bfw_u$ and $\bfv_i$ are in $[0,1]^D$, it is immediate to see that $x \in [0, 1]$. We sample $p_{u,a}$ from a uniform distribution $U(L_a, 1)$, where $L_a$ is a user-specified per-attribute parameter. In the special case when $L_a=1$ then $U(L_a, 1)$ becomes the linear utility. In our experiments, we set $L_a$ to 0.3 for sense-subjective tags and $L_a$ to 0.5 for the rest. 

\subsection{MovieLens-20m Data}
\label{app:movielens}
The MovieLens-20m data set consists of 20 million movie ratings on a scale from 1 to 5 and 465,000 free form tag applications for 27,000 movies by 138,000 users.
For our analysis, we transform all tags to lowercase and filter tag-and-rating data to only include the user-item-tags whose corresponding ratings are at least 4.
This results in around 235,000 user-item-tag triples, which comprise 20,068 unique tags. Many of these are applied to only a few movies or by a few users: 11,145 tags are only applied by a single user, and just 268 tags are applied to at least 50 unique movies. Since the CAVs are trained in 50-dimensional latent space, we restrict the CAV training data to the top 250 tags in terms of unique tagged movies. Inspection of the tag data further shows that tags that are applied by only a few users tend to be overly-specific or overly-generic rather than descriptive of the particular movies tagged.
For example, the tags `memasa's movies', `on dvr' and `bd-video' were applied by single users to 216, 210 and 192 movies, respectively. The tags `vhs' and `owned' were applied to 194 and 155 movies by 20 individual users.
To exclude these types of tags, we further filter the data to include only the top 250 tags in terms of unique users who have used the tag at least once. This leaves us with a total of 164 tags for evaluation.
User-item-tag triples are then split into train-test with a roughly $(0.75,0.25)$ split, with \emph{all examples} for any specific user-item pair present in exactly one of these subsets, i.e., if user $u$ rated item $i$, the all tags $g$ applied by $u$ to $i$ (if any) will only be in either train or test data set.

\subsection{More Details on Training Procedures}
\label{app:training}

We provide additional details on the training methods used in our experiments.

\subsubsection{Weighted Alternating Least Squares}
\label{app:wals_training}

We learn user and item embeddings using collaborative filtering. We consider two approaches, matrix factorization using \emph{weighted alternating least squares} (WALS) \cite{wals:icdm08}, which we describe in this subsection and
a two-tower DNN (next subsection).
Collaborative filtering typically generates a low-rank approximation of the user-item ratings matrix, which models both users and items in the low-dimensional (latent) feature space. Both users and items are represented by feature vectors in $X \subset\mathbb R^d$, where we let $\phi_U$ (resp., $\phi_I$) be the mapping from users (resp., items) to features.

In our experiments with linear utility models, we learn $(\phi_U, \phi_I)$ using 
WALS which uses the following regularized objective:
\begin{align}
(\phi^*_U, \phi^*_I)\in\arg\min\,\,\sum_{u,i}c_{u,i}(\hatr_{u,i}-r_{u,i})^2+ \kappa(||\phi_U||^2 + ||\phi_I||^2),
\end{align}
where $c_{u,i}$ is the confidence weight of the predicted rating $\hatr_{u,i}$ and $\kappa > 0$ is the regularization parameter. We pick the feature vectors $(\phi^*_U, \phi^*_I)$ based on the best validation loss and train with an item-oriented confidence weight, i.e., 
$c_{u,i}\propto m -\sum_u r_{u,i}$, which assigns lower weight to less-frequently rated or lower-rated items.
Example confidence weights include: (i) uniform, i.e., $c_{u,i}=\delta$ for some $\delta >0$ for missing entries $(u,i)$ and $c_{u,i}=1$ otherwise; (ii) user-oriented, i.e., $c_{u,i}\propto\sum_i r_{u,i}$ which assigns higher confidence to users with more ratings, assuming he/she has greater item familiarity; and (iii) item-oriented, i.e., 
$c_{u,i}\propto m -\sum_u r_{u,i}$, which assigns lower weight to less-frequently rated or lower-rated items. In our experiments with linear utility models, we train a collaborative filtering model $\Phi = (\phi_U, \phi_I)$ using \emph{WALS} given ratings $\bfR$, in order to obtain user and item feature vectors.
We set the regularization parameter $\kappa$ to 250 for MovieLens and 1 for the synthetic data. The number of WALS iterations is 100. We train with the third option for confidence scores and pick the feature vectors $(\phi^*_U, \phi^*_I)$ based on the best validation loss.

\subsubsection{DNN Training}
\label{app:DNN_training}
When training nonlinear CAVs (i.e., when user utility is nonlinear with respect to the soft attribute in question),
user and item embeddings $\phi_U$ and $\phi_I$ are generated by DNNs, $N_U$ and $N_I$, respectively, and (as above) predicted ratings are generated by taking dot products, using a ``two-tower'' architecture. following deep neural network (DNN) models. Letting $\theta_U, \theta_I$ denote the parameters of $N_U, N_I$,\footnote{We remove dependence of various terms (e.g., $\phi_U, \phi_I$, $\hatr_{u,i}$) on the DNN parameters $\theta_U, \theta_N$ to unclutter the notation when this dependence is obvious.}
similar to the linear case, we train the two-tower model by minimizing the regularized (squared) RMSE loss:.
\begin{align}
(\theta_U^*, \theta_I^*)\in\arg\min_{\theta_U,\theta_I}\,\,\text{RMSE}^2(\theta_U,\theta_I)+ \kappa(||\phi_U(\cdot;\theta_U)||^2 + ||\phi_I(\cdot;\theta_I)||^2)+ \rho(||\theta_U||^2 + ||\theta_I||^2),
\end{align}
where $\text{RMSE}(\theta_U,\theta_I) =\sqrt{\frac{1}{N}\sum_{u,i}(\hatr_{u,i}(\theta_U,\theta_I)-r_{u,i})^2}$, the predicted rating is $\hatr_{u,i}(\theta_U,\theta_I) = \phi^\top_{U}(u;\theta_U) \phi_{I}(i;\theta_I)$, $\kappa>0$ is the activity regularization constant, and $\rho>0$ is the weight regularization constant. Since the above objective function is nonlinear and nonconvex, we simply optimize this objective function with stochastic gradient descent based approaches, e.g., ADAM \cite{adam}.

We use one-hot encodings of users and items as inputs to $N_U$
and $N_I$, respectively. For simplicity, we use the same architecture for both the user DNN and item DNN, which is a feed-forward network with $\ell$ layers, in which the first layer has no bias term and has a dimension of $d$---we let $d$ have the same dimension as the linear embedding space (25 for synthetic data, 50 for MovieLens). The main motivation for this architectural choice is to allow weight initalization of the first layer using the linear user and item embeddings as a form of warm start. In our experiments, we use a DNN architecture with $\ell = 3$ fully connected layers, all of size (width) $d$, and ReLU activations. As in the linear case, we set the activity regularization parameter $\kappa = 250$ for MovieLens and $\kappa = 1$ for the synthetic data. For weight regularization, we set $\rho = 1$ in all the experiments. Given the trained DNNs, the activation vectors for training (item) CAVs are simply the $\ell$-th intermediate activation layer $N_{I,\ell}(i)$, extracted from the trained item DNN.
We treat the choice of intermediate layer $\ell$ as a tunable hyper-parameter chosen to optimize the downstream task (CAV prediction). 

\subsubsection{PITF Training}
\label{app:pitf}


When evaluating CAVs for their ability to predict tag usage, we use the \emph{PITF (pairwise interaction tensor factorization)} algorithm \cite{rendle2010} as a natural baseline, since it is a method designed to predict (or recommend) tags for users. We train PITF using tag triples $t_{u,i,t}$ with the same train-test split as in CAV training. Notice that positive tags are generally more sparse in all data sets, to balance the portion of positive and negative samples for PITF training we also use the same negative sampling methodology as in CAV training.
For fairer evaluation, we use the same embedding dimension as in CAV for PITF training (50 for MovieLens, 25 for synthetic data), but notice that the resulting PITF model is still much larger than the CAV model since PITF learns an embedding for each user, item, tag-user and tag-item, while CAV embedding is user-independent. In evaluating the accuracy of PITF, we do not adopt the F-score over Top-N items as in \cite{rendle2010}. but instead we check if $sign(y_{u, i, g}) = t_{u,i,g}$, where $y_{u, i, g}$ is the prediction of PITF. In addition, we evaluate with  \emph{Spearman rank correlation coefficient} which is translation-invariant.
We employ the implementation at \texttt{https://github.com/yamaguchiyuto/pitf/} to train the PITF embedding. For hyper-parameters, we set the learning rate ($\alpha$ in \cite{rendle2010}) to 0.001 for MovieLens and 0.0002 for synthetic data, the regularization parameter ($\lambda$ in \cite{rendle2010}) to 0.01 for MovieLens and 0.00005 for synthetic data. These parameters are tuned with a validation set. The number of PITF training iterations is set to 100.

\subsection{Additional Critiquing Details}
\label{app:critiquing}

We fill in a few additional details of the critiquing experiments in Section~\ref{sec:critiquing}. We emphasize again that our aim is not to evaluate state-of-the-art critiquing and elicitation methods, but to demonstrate the use of CAV-based soft-attribute semantics in allowing users to effectively interact with the recommender systems and refine the recommendation results.

In the synthetic data experiment, we construct each user's \emph{estimated ideal item} using the knowledge of her ground-truth utility function, which is either linear or single-peaked linear in each (latent or soft-attribute) dimension. We note that a user's (actual or estimated) `` ideal'' item may not actually exist in the item corpus $\calI$. Insisting that the user's ideal item exist is unrealistic, since it requires a dense item space. Moreover, the user generally does not know the identify of the actual best item in $\calI$, since this would assume too great a state of knowledge for most users. Instead, we assume each $u$ has a rough estimate of the maximum and minimum levels any tag/attribute can attain in the item corpus and uses this to drive her critiques. (For example, we truncate sampled items to $[0,1]^D$.)
Note that when user utility is linear, the ideal item must occur at the boundary of item space, so the estimates inform her estimated ideal.
In the MovieLens experiment, we use the ratings data to derive an estimated ground-truth utility for each \emph{test user} (who must have rated sufficiently many items as described in the main text). 

During the critiquing process, the recommender system updates its user embedding based on the user's response.\footnote{We use the average of all learned user embeddings as the systems's prior.}
We assume item embeddings $\phi_I(i)$ are fixed, and use a \emph{simple heuristic recommendation strategy} for incorporating critiques.\footnote{We emphasize that the recommendation strategy used and method for incorporating critiques are fairly generic and are not intended to reflect the state-of-the-art, since our goal is to measure the ability to exploit learned CAVs. More elaborate strategies for updating user embedding are possible, including the use of Bayesian updates relative to a prior over the user embedding \cite{vendrov2020gradient}. Here instead we adopt a simple heuristic, based on \cite{luo2020latent}, to focus attention on the CAV semantics itself.} Given a user embedding $\phi_U(u)$, the recommender system scores all items $i$ in the corpus with respect to utility $r_{i,u} = \phi_I(i)^T\phi_U(u)$, and presents the slate $S$ of the $k$ top scoring items. 

Suppose at step $t>0$ the user critiques $S$ with a specific tag $g$ (and a specific direction, \emph{more} or \emph{less}). The recommender system updates the user embedding in response to this critique using a simple heuristic update function: $\phi_U(u) \leftarrow \phi_U(u) + \text{Sgn}\cdot\alpha_t(g) \cdot \phi_g$. Here $\text{Sgn}\in\{+1, -1\}$ indicates the direction of the move ($+1$ more,  $-1$ less), and $\alpha_t(g)$ is a step size that controls the scale of the move of the user embedding in the direction of tag $g$'s CAV. For each $g$, we decay the step size $\alpha_t(g)$ with each critique using $g$, $\alpha_t(g) = \alpha_0(g) / (1 + t)$, where $\alpha_0$ is $g$'s initial step size. This ensures that the updating process converges to a stable point (and does not cycle or repeat slates of items) given the coarse control mechanism offered to the user. 
If the tag $g$ is sense-subjective, the critique is interpreted relative to the recommender system's estimate of $u$'s sense (or user cluster) based on past usage.

In our experiments, $\alpha_0(g)$ is constant ($\alpha_0$) across all tags, and we treat it as a tunable hyper-parameter selected to optimize user utility metrics such as $\UMU$ and $\UAU$ utilities.\footnote{Ultimately, this heuristic adjustment should be tuned to the specifics of a real-world user response models.} 
The critiquing results we report are based on the set of hyper-parameters optimized using a validation set. 

\end{document}